\documentclass[journal=jctcce,manuscript=article]{achemso}

\usepackage{mathtools,amssymb,graphicx}
\usepackage[plainpages=false,pdfpagelabels,colorlinks=true,linkcolor=red,urlcolor=blue,citecolor=blue,pdftitle={Title},pdfauthor={},pdfdisplaydoctitle=true,pdfduplex=DuplexFlipLongEdge]{hyperref}
\usepackage{soul} % to use \st for strikeout 
\usepackage[dvipsnames,usenames]{color}
\usepackage[normalem]{ulem}
\usepackage{tabularx}
\usepackage{graphicx}
\usepackage{xcolor}
\usepackage{bbold, verbatim} % to generate identity 1
\usepackage{float}
\graphicspath{ {./figures/} }
\usepackage{tabularx}
\usepackage{rotating}
\usepackage{braket}
\usepackage{multirow}
\usepackage{ulem}
\usepackage{amsmath}

\newcommand{\tvb}{\textsc{TurboRVB}} 
 
\newcommand{\tbg}{\textsc{TurboGenius}} 
\newcommand{\pyscf}{\textsc{PySCF}} 
\newcommand{\trexio}{\textsc{TREX-IO}}

\makeatletter
\def\Hline{
\noalign{\ifnum0=`}\fi\hrule \@height 1pt \futurelet
\reserved@a\@xhline}
\makeatother

\author{Kousuke Nakano}
\email{kousuke_1123@icloud.com}
\affiliation[NIMS]
{Center for Basic Research on Materials, National Institute for Materials Science (NIMS), Tsukuba, Ibaraki 305-0047, Japan}
\alsoaffiliation[SISSA]
{International School for Advanced Studies (SISSA), Via Bonomea 265, 34136, Trieste, Italy}

\author{Sandro Sorella}
\affiliation[SISSA]
{International School for Advanced Studies (SISSA), Via Bonomea 265, 34136, Trieste, Italy}

\author{Dario Alf{\`e}}
\affiliation[UNiNa]{Dipartimento di Fisica Ettore Pancini, Universit\`a di Napoli Federico II, Monte S. Angelo, I-80126 Napoli, Italy}
\alsoaffiliation[UCL]{Department of Earth Sciences, University College London, Gower Street, London WC1E 6BT, United Kingdom}
\alsoaffiliation[TYC]{Thomas Young Centre and London Centre for Nanotechnology, 17-19 Gordon Street, London WC1H 0AH, United Kingdom}

\author{Andrea Zen}
\email{andrea.zen@unina.it}
\affiliation[UNiNa]{Dipartimento di Fisica Ettore Pancini, Universit\`a di Napoli Federico II, Monte S. Angelo, I-80126 Napoli, Italy}
\alsoaffiliation[UCL]{Department of Earth Sciences, University College London, Gower Street, London WC1E 6BT, United Kingdom}

\title{Beyond single-reference fixed-node approximation in {\emph {ab initio}} Diffusion Monte Carlo using antisymmetrized geminal power applied to systems with hundreds of electrons} 

\begin{document}

\begin{tocentry}
%\centering
 \includegraphics[width=7.0cm]{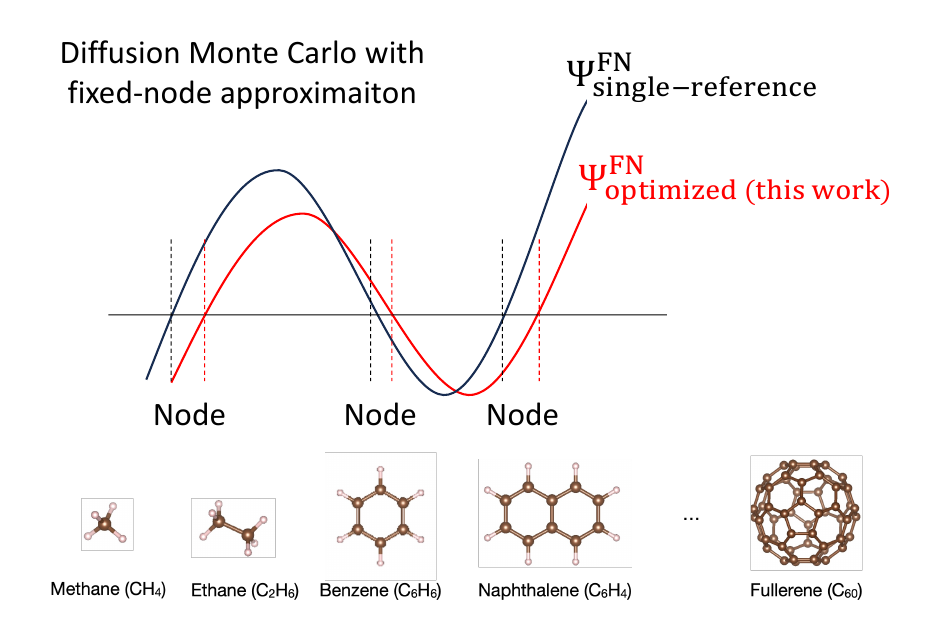}
 \label{For Table of Contents Only}
%Some journals require a graphical entry for the Table of Contents. This should be laid out ``print ready'' so that the sizing of the text is correct. Inside the \texttt{tocentry} environment, the font used is Helvetica 8\,pt, as required by \emph{Journal of the American Chemical Society}. The surrounding frame is 9\,cm by 3.5\,cm, which is the maximum permitted for  \emph{Journal of the American Chemical Society} graphical table of content entries. The box will not resize if the content is too big: instead it will overflow the edge of the box. This box and the associated title will always be printed on a separate page at the end of the document.

\end{tocentry}

\begin{abstract}
Diffusion Monte Carlo (DMC) is an exact technique to project out the ground state (GS) of a Hamiltonian. 
Since the GS is always bosonic, in fermionic systems the projection needs to be carried out while imposing anti-symmetric constraints, which is a nondeterministic polynomial hard problem. 
In practice, therefore, the application of DMC on electronic structure problems is made by employing the fixed-node (FN) approximation, consisting of performing DMC with the constraint of having a fixed predefined nodal surface. 
How do we get the nodal surface?
The typical approach, applied in systems having up to hundreds, or even thousands of electrons, is to obtain the nodal surface from a preliminary mean-field approach (typically, a density functional theory calculation) used to obtain a single Slater determinant. This is known as {\emph{single reference}}.
In this paper, we propose a new approach, applicable to systems as large as the C$_{60}$ fullerene, which improves the nodes by going beyond the single reference. 
In practice, we employ an implicitly multireference ansatz (Antisymmetrized Geminal power wavefunction constraint with molecular orbitals), initialized on the preliminary mean-field approach, which is relaxed by optimizing a few parameters of the wave function determining the nodal surface by minimizing the FN-DMC energy.
We highlight the improvements of the proposed approach over the standard single reference method on several examples and, where feasible, the computational gain over the standard multireference ansatz, which makes the methods applicable to large systems. We also show that physical properties relying on relative energies, such as binding energies, are affordable and reliable within the proposed scheme. 
\end{abstract}

%%%%%%%%%%%%%%%%%%%%%%%%%%%%%%%%%%%%%%%%%%%%%%%%%%%%%%%%%%%%%%%%%%%%
%% main body
%%%%%%%%%%%%%%%%%%%%%%%%%%%%%%%%%%%%%%%%%%%%%%%%%%%%%%%%%%%%%%%%%%%%

\section{Introduction}
% A very general introduction of electronic structure calculation
%\paragraph{{\textit{Introduction}} $-$}
{\emph{Ab initio}} electronic structure calculations, which compute the electronic structure of materials non-empirically, have become an essential methodology in the materials science and condensed matter physics communities. Density functional theory (DFT), a mean field approach which was originally proposed by Kohn and Hohenberg~{\cite{1964HOH}}, is the most widely used methodology for {\emph{ab initio}} electronic structure calculations. 
DFT has enjoyed widespread success, despite its reliance on the so-called exchange-correlation (XC) functional, whose exact form is yet to be discovered.
Although many XCs have been proposed, no functional that performs universally well for all materials is established. 

Several methodologies transcend the mean-field paradigm. 
% CCSD(T)
For example, in the quantum chemistry community, the coupled cluster method with single, double, and perturbative triple excitations \cite{Bartlett2007}, denoted as CCSD(T), is widely recognized as the gold-standard approach, balancing  accuracy and computational efficiency. This technique has been employed as a reference in many benchmark tests, both for isolated and periodic systems~\cite{Bartlett2007, 2013_gold_CCSD_T, Rezac2016, Al-Hamdani2019}. 
CCSD(T) is mostly applied in relatively small systems, as it becomes very computationally intensive as the simulated systems get larger (hundreds of electrons or more).
Moreover, despite the many successes of CCSD(T), there are a few problems where CCSD(T) fails, mostly attributed to the multireference character of a chemical system (strong correlation) and where other methods, more expensive computationally, are needed \cite{Szalay2012}.
% DMC
A different approach, adopted by the condensed matter community as the gold standard, is the diffusion Monte Carlo (DMC) method~{\cite{2001FOU}}.
DMC has good scaling with the system size and it uses algorithms that can be parallelized with little or no efficiency lost, fully exploiting modern supercomputers and making relatively large systems treatable.

CCSD(T) and DMC predictions typically show consensus in the computed physical properties, such as heats of formation and binding energies, and good agreement with experiments \cite{Dubecky2016, Al-Hamdani2017, Brandenburg2019_W@Gr, Zen2018_PNAS, Al-Hamdani2019, 2021YAS, Shi2023}.  It was believed that CCSD(T) and DMC would also agree on extended systems, but recent findings by Al-Hamdani et al.~{\cite{2021YAS}} have unveiled discrepancies in binding energy calculations between these methods for large systems, such as a C$_{60}$ buckyball inside a [6]-cycloparaphenyleneacetylene ring (C${60}$@[6]CPPA). 
It is unclear which approach is to be trusted in these tricky cases.
These findings raise a pivotal question: 
%Which is the more reliable approach for non-covalent interactions between large systems? 
what is the reference approach for non-covalent interactions between large systems?
To answer this question, Al-Hamdani et al. discussed possible discrepancy sources coming from uncontrollable errors existing in both CCSD(T) and DMC. 
Both approaches employ some approximations and have their weaknesses, and the debate is still open. 
To draw a more conclusive determination, one should develop a scheme which mitigates the impact of uncontrollable errors in the methods.
In this work, we focus on improvements in the DMC approach that alleviate its largest source of error: the fixed-node (FN) approximation.

%{\vspace{2mm}}
%The important sampling implementation of DMC realises a projection to the ground state from a given trial wavefunction.
DMC yields the exact ground state in bosonic systems. 
In fermionic systems (for instance, in electronic structure calculations) DMC suffers from the so-called negative sign problem, arising from the fact that the fermionic ground state has positive and negative regions.
The negative sign problem in the DMC method for fermions has been proven to be a nondeterministic polynomial hard problem~{\cite{2005TRO}}; thus, it seems unrealistic to find a general solution at present. 
This problem is avoided, in practice, by modifying the DMC projection algorithm  with the introduction of the fixed-node (FN) approximation, where the projected wave function $\Phi_\text{FN}$ is constrained to have the nodes of a predetermined guiding function $\Psi_{\text{T}}$.
The FN approximation keeps the projected wave function $\Phi_\text{FN}$ antisymmetric, but $\Phi_\text{FN}$ is the exact ground state $\Phi_0$ only if its nodes are exact. 
A general property of $\Phi_{\rm FN}$ is that it is always the closest function to $\Phi_0$ within the FN constraint.
For trial functions obtained from mean field approaches, such as Hartree-Fock(HF) or DFT, it is generally believed that the error associated to the FN approximation is small and  benefits from a large error cancellation in the evaluation of binding energies~{\cite{Dubecky2016}}. 
However, the FN error is typically not accessible, as $\Phi_0$ is unknown, and this yields an uncontrollable error in FN-DMC.

In standard FN-DMC simulations, the nodal surface is given by an approximate wave function, which is typically obtained starting from a mean-field approach, such as HF or DFT. 
The variational principle can still apply to FN-DMC, \footnote{In all-electron calculations, FN-DMC is always variational, meaning that the lowest FN energy $E_\text{FN}$ is obtained when the exact nodal surface is used, otherwise $E_\text{FN}>E_0$. When pseudopotentials are employed there are also non-local operators in the hamiltonian. This yields to a problem similar to the sign-problem, which requires a further approximation. There are a few alternatives to deal with pseudopotentials in DMC: 
the locality approximation (LA) \cite{LA:mitas91}, the T-move (TM) \cite{Casula06, Casula10}, the determinant locality approximation (DLA) and the determinant locality T-move (DLTM)  \cite{2019ZEN}. TM and DLTM are variational, meaning that their energy ($E_\text{FN,TM}$ or $E_\text{FN,DLTM}$) is an upper bond of the exact ground state energy $E_0$.
%and the exact FN-DMC solution is achieved with the employment of a trial wave function to project the non-local operators which is exact in proximity of the nuclei, other than having exact nodes.
} and so to go beyond the mean-field solution, one should optimize the given nodal surface by minimizing the the FN-DMC energy $E_\text{FN}$ (which is the expectation value of $\Phi_\text{FN}$), going in the direction of the exact wavefunction $\Phi_0$ and the exact energy $E_0$. 
This procedure is seldomly followed in DMC simulations, especially on large systems (say, with hundreds or thousands of electrons), as it is hardly affordable computationally and the uncertainty on the optimization of the FN surface could be easily comparable, if not larger, than the binding energy under consideration. Thus, the standard approach is to just keep the nodal surface of the Slater determinant built with the Kohn-Sham orbitals obtained from a DFT calculation.
% works well 
Whilst the FN surface from DFT might be suboptimal, this approach typically yields quite reliable results, especially in the evaluations of non-covalent interactions, due to very favourable error cancellations~\cite{Dubecky2016, Al-Hamdani2019}. 

% Improve nodes
In smaller systems (with say, tens of electrons), it is possible to improve the nodal surface, and the most standard approach is to use an ansatz that has more degrees of freedom than the initial Slater determinant, such as the antisymmetrized geminal power (AGP)~\cite{2003CAS, 2004CAS, 2009MAR}, the Pfaffian~\cite{Bajdich2006, Bajdich2008, Genovese2020}, the complete active space~\cite{Toulouse2008, CAS_Zimmerman2009}, the valence bond~\cite{GVB_Anderson2010, Braida2011}, the backflow~\cite{2006DRU, 2006RIO, Bajdich2008}, and multideterminant expansions~\cite{Toulouse2007, Seth2011, Clark2011, Morales2012, Filippi2016, Scemama2016, Dash2018, Scemama2020}, including methods employing neural networks and machine learning techniques~\cite{FermiNet_Pfau2020, Li2022, Ren2023, PauliNet_Hermann2020, Choo2020, ForwardLaplacian_Li2023, Psiformer_VonGlehn2022}. 
The standard approach here is to optimize the wave function parameters at the level of theory of variational Monte Carlo (VMC)~{\cite{2003CAS, 2018SOR, 2019NAK, 2020GEN, 2022AMM, 2023RAG}}.
Indeed, optimization at the FN-DMC (FN-opt) level implies further difficulties, as we will discuss below.
However, optimization at the VMC (VMC-opt) level has some flaws.
% Issues of VMC-opt for DMC
%The main issue is that the nodal surface obtained using VMC gradients does not guarantee a better nodal surface because the variational principle on the nodal surface only applies to DMC, and not, in general, to VMC \tobe{\bf is this true?}.
% Riscrivo
In VMC-opt the object that is optimized is the variational wave function $\Psi_{\text{T}}$, which is obtained from the product of one of the ansatze discussed above and the Jastrow factor~\footnote{
The Jastrow factors correlates explicitly the electrons, its a symmetric positive function, so it recovers dynamical correlation and it does not change the nodal surface. 
}.
%, by minimising either the VMC energy or the VMC variance (based on the variational principle or on the zero-variance property). 
The closer $\Psi_{\text{T}}$ gets to the ground state $\Phi_0$, the smaller its VMC energy (variational principle) and its VMC variance (zero-variance property) are. 
VMC-opt explores the parameters variational space, seeking the set which minimises the VMC energy or  the VMC variance, and it is often done by employing the VMC gradient.
It is not guaranteed that the parameters obtained from VMC-opt are those giving the best possible nodal surface allowed by the employed ansatz (unless we are in the limit case where $\Psi_{\text{T}}$ yields VMC with zero variance, such that we know that $\Psi_{\text{T}}$ is an eigenstate of the Hamiltonian).
Although this approach, in practice, gives a better nodal surface than the DFT one, it sometimes gives unreasonable outcomes, e.g., it overestimates binding energies, as revealed in this work. 
%
% FN-opt
It would be desirable, instead, to implement an optimization at the FN-DMC level of theory, where the parameters of the function $\Psi_{\text{T}}$ giving the nodal surface are optimized so as to minimize the FN energy. 
This would guarantee to find the best nodal surface allowed by the adopted wave function ansatz.
To the best of our knowledge, the first attempt to directly optimize the variational parameters included in a trial WF at the FN-DMC level was done by Reboredo et al.~{\cite{2009REB}} in the {\emph{ab initio}} framework. They proposed a way to iteratively generate new trial wavefunctions to get a better nodal surface. They generalized the method to excited states~{\cite{2009REB2}} and finite temperatures~{\cite{2014REB}} and also applied for large systems such as C$_{20}$~{\cite{2010BAJ}}. Very recently, McFarland and Monousakis~{\cite{2022MCF}} reported successful energy minimizations with approximated and exact FN gradients. They proposed to optimize nodes using a combination of FN gradients and the projected gradient descent (PGD) method. The PGD method works for Be, Li$_2$, and Ne using all-electron DMC calculations~{\cite{2022MCF}}, while it has been successful only for small molecules.

%{\vspace{2mm}}
When it comes to optimizing the nodal surface of a large system, the main problem is that the number of variational parameters determining the nodal surface often scales more than linearly with the size of the system. For instance, the number of variational parameters in the determinant part of the AGP ansatz scales with $O(L^2)$, where $L$ is the number of basis functions in a system. It makes the parameter space to be optimized so complex that the optimization is easily trapped in local minima and one cannot find the true ground state. 
%Moreover, the uncertainty in the optimization of the parameters introduces fluctuations in the evaluation of the energy, let's call it the {\em optimization uncertainty}, which increases with the system size and with the number of variational parameters, and can be reduced only at the cost of increasing the statistical sampling (and the computational cost). 
Moreover, since the optimization algorithms are stochastic, there is always an additional uncertainty on the optimized parameters, which are not going to be exact and the corresponding QMC energy has therefore an optimization bias.
The optimization bias increases with the system size and with the number of variational parameters, and can be reduced only at the cost of increasing the statistical sampling (and the computational cost).
The evaluation of a binding energy implies the difference between two or more DMC energies, and it is often a tiny fraction of the total energy. 
Therefore the optimization uncertainty can often be comparable to the binding energy, making the evaluation of the interaction energy unreliable. 
Moreover, we need to verify that the adopted approach satisfies basic physical properties, such as being size-consistent~\footnote{An approach is size consistent if the energy of a system constituted by two or more non-interacting subsystems (e.g., two molecules far away) is the same of the sum of the energies of the subsystems.}.
At the VMC level, the size-consistency is a property of the wave function ansatz employed, and it depends on the optimization procedure. At the FN-DMC level, size consistency might depend on some choices on the algorithm~\cite{2015ZEN}, on the ansatz of the wave function providing the FN constraint, and on the optimization.

%\tobe{XXX Summarise main points on new scheme: systematically better than JSD, size consistent (non-covalent interacions), large systems are doable. XXX}
%{\vspace{2mm}}
In this paper, we propose a scheme which aims to address these issues. In particular, our scheme satisfies the following points: (i) it is systematically more accurate than the standard approach of employing a single Slater determinant, (ii) it is size consistent, and (iii) it is applicable also to large systems.
%
%In our scheme, we optimize the nodal surface using the FN gradients. 
The idea underlying the present work is the combination of the AGP wavefunction consisting of molecular orbitals (AGPn)~{\cite{2009MAR}},
the use of natural orbitals, 
and the optimization of its nodal surface using FN gradients on a selected subset of the AGPn parameters. 
In particular, we initialize the orbitals in the AGPn wave function using natural orbitals, which are kept fixed afterwords, such that only the coefficients combining them are optimized to relax the nodal surface.
We call this scheme the fixed node antisymmetrized geminal power active-space (FNAGPAS). 
Since the orbitals are fixed, this results in a much smaller number of variational parameters in the ansatz; thus, one can apply it for larger systems, such as C$_{60}$ fullerene. We show that our scheme gives a better nodal surface (i.e., a lower energy in the FN-DMC calculation) compared to the typical Slater-Jastrow ansatz, and it reliably describes also strongly correlated systems (such as diradicals). We show that the use of FN-opt is important to fulfill the size-consistency property. %, which is not necessarily fulfilled using VMC-opt. 
%We found that physical properties relying on relative energies, such as binding energy and torsion energy of molecules, are also improved by our proposed scheme, compared with those obtained by the standard procedure.

\section{The FNAGPAS scheme}
\label{sec:methods}

\begin{figure*}[htbp]
\centering
\includegraphics[width=1.0\textwidth]{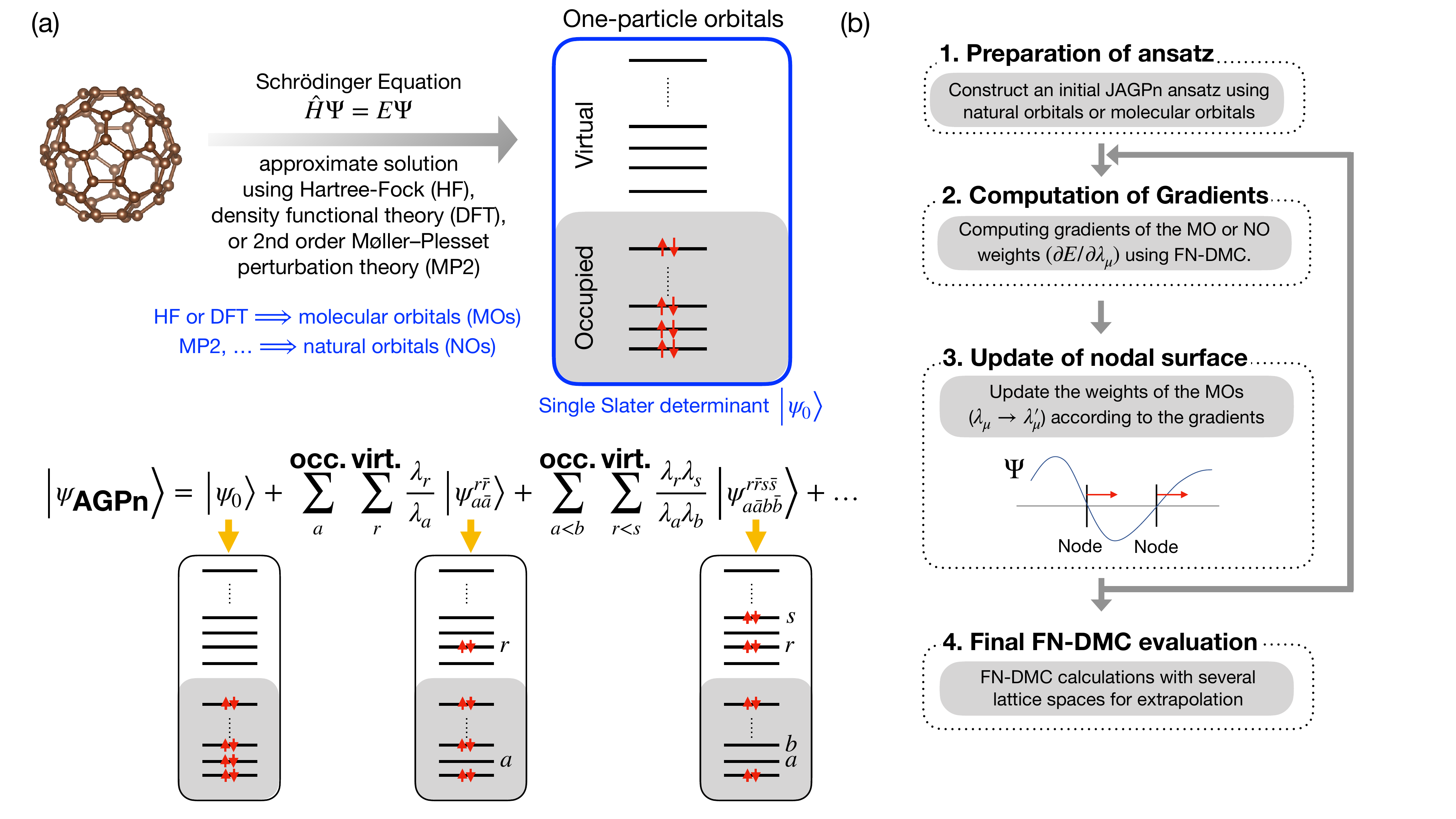}
\caption{
Panel a: Schematic illustration of the FNAGPAS scheme. We perform a preliminary mean field calculation to obtain molecular orbitals (MOs), followed by a correlated calculation yielding natural orbitals (NOs). The AGPn ansatz correspond to a multideterminant expansion built on the NOs and depending on the coefficients $\lambda_i$ associated to each orbital $i$ and optimized in order to minimise the FN energy.
Panel b: Flowchart illustrating the FNAGPAS scheme workflow. 
}
  \label{fig:concept}
\end{figure*}

We describe here the scheme that we suggest to improve the accuracy of FN-DMC over the traditional single determinant Slater-Jastrow ansatz. 
The key idea is the combination of the AGP wavefunction constraint with molecular orbitals (AGPn)~{~{\cite{2009MAR}}} and the optimization of the ansatz using approximated FN gradients~\cite{2022MCF}. 
We describe the ansatz in the following section, assuming an unpolarized system for simplicity. The schematic illustration explaining the key concept and its workflow is shown in figure~{\ref{fig:concept}}.

%%%%%%%%%%%%%%%%%%%%%%%%%%%%%%%%%%%%%%%%%%%%%%%%%%%%
%%%%%%%%%%%%%%%%%%%%%%%%%%%%%%%%%%%%%%%%%%%%%%%%%%%%

%\subsection{The wave function ansatz}
%\vspace{2mm}
The real-space quantum Monte Carlo typically employs a many-body WF ansatz $\Psi$ written as the product of two terms, $\Psi_\text{QMC}  =  \Phi _\text{AS} \times \exp J$. The term $\exp J$, conventionally dubbed Jastrow factor, is symmetric, and the term $\Phi _\text{AS}$ is antisymmetric. The Jastrow factor is explicitly dependent on electron-electron distance, and often includes electron-nucleus and electron-electron-nucleus terms.\footnote{See Ref.~{\citenum{2020NAK}} for details about the functional form of terms implemented in the TurboRVB package used for this work.} 
The nodal surface of a WF is determined by the antisymmetric part $\Phi _\text{AS}$ (because $\exp J \ge 0$).
Thus, in FN-DMC the accuracy of the results depends crucially on the quality of the nodes of $\Phi _\text{AS}$.

% the JSD function
The antisymmetric part of a trial WF is initially constructed from a mean-field self-consistent-field (SCF) approach, such as DFT or HF.
The standard QMC setup in large systems is to define $\Phi _\text{AS}$ as the single Slater determinant (SD) obtained from such preliminary SCF calculation. 
The corresponding $\Psi_\text{QMC}$ is dubbed JSD.
Therefore, the nodes of JSD are predefined before any QMC calculation and unrelaxed. 
Initializing the SD using different setups for the SCF calculations (e.g., different exchange correlation functionals) leads to slightly different total energies, but most of the times, the interaction energies (which are evaluated from energy differences between two or more systems) are almost unaffected by the details of the preliminary SCF calculation, expecially for weak non-covalent interactions. This is an indication that there is an almost perfect cancellation of the error induced by the FN approximation within the JSD ansatz, provided that the SD is initialized consistently in all systems.

% Beyond JSD, are Eb changed?
However, changing the setup of the SCF calculation only allows the nodes to move within the variational freedom of a single Slater determinant.
By contrast, giving $\Phi _\text{AS}$ the variational freedom to relax the nodes beyond the JSD ansatz leads to an improvement of the FN-DMC total energy of the system \cite{Annaberdiyev2020}, and possibly also the interaction energies could change.
The challenge that we take here is to generalize the ansatz in a way that large systems are still doable.

Here we suggest to use the AGP ansatz as $\Phi _\text{AS}$.
AGP is an implicitly multideterminant ansatz \cite{2014ZEN,2015ZEN}, which corresponds to a constrained zero-seniority expansion, as illustrated schematically in figure~\ref{fig:concept}. 
The evaluation of an AGP function can be reduced to the computation of a determinant, therefore the AGP ansatz is computationally comparable to a Slater determinant (differently from explicitly multideterminant functions), thus ensuring the cubic scaling with the system size of both the variational and FN algorithms~\footnote{ It is generally claimed that the cost of FN-DMC scales as the cube of the number of electrons $N_{\rm{el}}$. This is true for simulations where the antisymmetric part of the wave function can be computed as a determinant and $N_{\rm{el}}$ up to roughly a thousand. For larger systems the cost for a monte Carlo step is ${\cal O}(N_{\rm{el}}^3)$ and therefore the cost of FN-DMC is quartic. }.
% definition of AGP and AGPn
The AGP ansatz for a system of $N_{\rm{el}}$ electrons is 
\begin{equation}\label{eq:AGP}
\Psi_{\rm AGP} = \hat{\cal A} [
g({\bf x}_1,{\bf x}_2)
g({\bf x}_3,{\bf x}_4)
\ldots
g({\bf x}_{N_{\rm{el}}-1},{\bf x}_{N_{\rm{el}}})]
\end{equation}
(we are assuming for simplicity an unpolarized system with even numbers of electrons, but the ansatz can be generalized as discussed in Ref.~\citenum{2003CAS}), 
where $\hat{\cal A}$ is the antisymmetrization operator and the function $g$ is the geminal function
$g({\bf x}_1,{\bf x}_2) = f({\bf r}_1, {\bf r}_2) {\alpha(1)\beta(2)-\beta(1)\alpha(2) \over \sqrt{2}}$,
which is a paring function between two electrons with coordinates ${\bf x}_1$ and ${\bf x}_2$ forming a spin singlet. 
The spatial part $f({\bf r}_1, {\bf r}_2)$ is symmetric, and it can be written in terms of a basis set $\left\{ \chi_\mu \right\}$ for the single electron orbital space as follows:
\begin{equation}\label{eq:f}
f({\bf r}_1, {\bf r}_2) = 
%\sum_{\mu,\nu}^{L,L} 
\sum_{\mu}^{L} 
\sum_{\nu}^{L} 
c_{\mu\nu} \, \chi_\mu \,({\bf r}_1) \chi_\nu({\bf r}_2)
\end{equation}
where $\mu$ and $\nu$ runs over all the $L$ basis orbitals, and $c_{\mu\nu}$ are variational parameters.
Notice that in general $L\gg N$, and the number of variational parameters $c_{\mu\nu}$ is equal to $L^2$.
The parameters define a $L\times L$ symmetric matrix $\bf C$ (the symmetry of $f$ implies $c_{\mu\nu}=c_{\nu\mu}$), so there is an orthogonal transformation $\bf U$ which diagonalizes $\bf C$ and allows to rewrite $f$ as:
\begin{equation}\label{eq:f2}
f({\bf r}_1, {\bf r}_2) = \sum_{\mu}^{L} \lambda_{\mu} \, \phi_\mu \,({\bf r}_1) \phi_\mu({\bf r}_2)
\end{equation}
where $\phi_\mu = \sum_\nu U_{\mu\nu} \chi_\nu$.
With no loss of generality we can assume that $\lambda$'s are ranked in decreasing order of their absolute value.
Notice that if only the first $N_{\rm{el}}/2$ $\lambda$'s are different from zero then $\Psi_{\rm AGP}$ corresponds to a single Slater determinant built on the orbitals $\phi_1,\ldots,\phi_{N_{\rm{el}}/2}$ occupied with both spin up and spin down electrons.
Since such Slater determinant built on orbitals from an SCF calculation is the standard QMC setup, and it typically delivers good results, we tried to relax the nodes by considering a subset $n_{\rm{orb}}$ (larger than $N_{\rm{el}}/2$ but $\ll L$) of the orbitals obtained from the SCF calculation. This is what we call the AGPn ansatz.

% Consideration about the Jastrow and its optimization
For an efficient and effective use in QMC the AGP and AGPn functions shall be multiplied by a Jastrow factor, yielding the so-called JAGP and JAGPn functions.
The Jastrow factor can have the same variational form used also in JSD, which allows for the JSD, JAGP and JAGPn functions to satisfy the cusp conditions and to effectively recover the dynamical correlations. 
Indeed, the main improvement of JAGP and JAGPn over JSD is their ability to capture static correlations, yielding to qualitatively different results on systems with an underlying multireference character, both at the variational and at the FN level of theory \cite{2014ZEN,2015ZEN}.
The optimization of the parameters in the Jastrow is usually quite a feasible problem also on large systems, as their number does not grow uncontrollably with the size of the system. 
In practice, every QMC code implements a slightly different functional form of the Jastrow, but their share the general features mentioned above.
The QMC code used in this work is TurboRVB~\cite{2020NAK}, an open-source package.
In TurboRVB the implemented the Jastrow factor (described in Ref. \citenum{2020NAK}) has a number of parameters growing linearly with the size of the system (as shown in the results section).

% The Idea
In this work, we keep the orbital frozen and optimize the coefficients $\lambda_1,\ldots,\lambda_{n_{\rm{orb}}}$ of the JAGPn ansatz using  FN-DMC gradients. 
A similar idea, but at the variational level, was also mentioned in a seminal work by Casula and Sorella to decribe the BCS paring function in iron-based superconductors~{\cite{2013CAS}}.
JAGPn dramatically reduces the number of variational parameters with respect to the JAGP ansatz, such that the optimization of the JAGPn function is doable even in pretty large systems, in contrast to JAGP which is affordable only on relatively small systems.
Nevertheless, employing JAGPn significantly improves the FN-DMC energy (as well as the variational QMC energy) over the results within the traditional JSD function, as we will show on the results section.
Of course, the JAGP ansatz has higher variational freedom than JAGPn, so JAGP can in principle improve further over JAGPn. However, in practice, we observe that FN-DMC energies obtained from the JAGP ansatz are comparable to those obtained from JAGPn on small systems (and both JAGP and JAGPn are significantly better that JSD), while, in large systems, JAGP is unaffordable because the optimization can be stuck at local minima at the variational level and can become unstable at the FN level. The latter instability is probably due to insufficient signal-to-noise ratios~{\cite{2017BEC}} that the QMC optimization always suffers from, but the origin of the instability is yet unclear. On intermediate systems, we notice that JAGP FN-DMC energy is worse than the JAGPn FN-DMC energy, as a clear indication that despite the higher variational freedom on JAGP, the optimization of that many parameters is not converging and there is too much noise on the parameters.

% Size consistency
The main problem of the AGP ansatz (and AGPn) is that it is not size-consistent at the variational level of theory, but JAGP (JAGPn) is size consistent if we employ a very flexible Jastrow factor \cite{SorellaRocca2007, Neuscamman2012}.
Since the FN-DMC corresponds to applying an infinitely flexible Jastrow factor to the determinant part, optimizing the AGPn parameters at the FN level ensures the size-consistency of our approach.

% Natural orbitals
A crucial point to make JAGPn almost as accurate as JAGP, despite employing only a small number $n_{\rm{orb}}$ of parameters $\lambda$'s, is to carefully choose the orbitals. We notice that the virtual orbitals obtained from SCF calculations are typically not optimal, as we need a large number of them (of the order of $L$) to converge to the best JAGPn FN energy. 
% (which is almost the same of the JAGP FN energy, on systems where JAGP is doable and the optimization reliable) {\bf Dario: I don't understand this statement. If we take the SCF orbtals for AGPn, aren't we just doing what we would do with JAGP?}. 
Moreover, if we cannot afford a systematic test of the convergence of $n_{\rm{orb}}$ for each system of interest, it is difficult to define a sensible criterion to decide which $n_{\rm{orb}}$ to pick. 
We solved both the problems by employing Natural Orbitals (NOs) for expanding the paring function, instead of using MOs. NOs were constructed from second-order M{\o}ller–Plesset (MP2) calculations. This is because the MP2 unoccupied orbitals incorporate perturbation effects and are physically better than those obtained with HF or DFT \cite{Gruneis2011}, as shown in the Supplemental Information. More specifically, we constructed natural orbitals by diagonalizing the density matrix obtained by MP2 calculations. We also notice that a method to construct NOs should be affordable also for large systems. This is also a reason why we chose MP2 for constructing NOs in this study.  In practice, from the weight of the NOs we can easily define a cutoff value to select $n_{\rm{orb}}$ on each system, and we notice that we get to converged results already with a value of $n$ that is not much larger than $N_{\rm{el}}/2$ ($n_{\rm{orb}}=N_{\rm{el}}/2$ would correspond to a single SD).

\section{Computational details}
%\vspace{2mm}
We applied our scheme to planar and twisted ethylenes, eight hydrocarbons (CH$_4$, C$_2$H$_4$, C$_2$H$_6$, C$_6$H$_6$, C$_{10}$H$_8$, C$_{14}$H$_{10}$, C$_{18}$H$_{12}$, C$_{20}$H$_{10}$), the C$_{60}$ fullerene, and water-methane dimer (see supplementary materials for their coordinates). 
The number of valence electrons treated in this study are 12, 12, 8, 12, 14, 30, 48, 66, 84, 90, 240, and 16, respectively.
The MP2 calculations (HF and DFT calculations for comparison) to generate nodal surfaces of trial wavefunctions were performed using \pyscf\ v.2.0.1~{\cite{2018SUN, 2020SUN}}. The trial wavefunctions were converted to the \tvb\ wavefunction format using \tbg~{\cite{2023NAK}} via \trexio~{\cite{2023POS}} files.  We employed the cc-pVQZ basis set accompanied with the ccECP pseudopotentials~{\cite{2017BEN}} for the eight hydrocarbons and C$_{60}$ fullerene, while the cc-pVTZ basis set accompanied with the ccECP pseudopotentials~{\cite{2017BEN}} for the water-methane and for the torsion calculation of ethylene. We employed [3$s$], [3$s$1$p$], and [3$s$1$p$] primitive Jastrow basis for H, C, and O atoms, respectively. The Jastrow factor and the weights of the natural orbitals in the paring function (i.e., the nodal surface of a WF) were optimized using the stochastic reconfiguration method~{\cite{1998SOR}} implemented in \tvb~{\cite{2020NAK}} with an adaptive hyperparameter~{\cite{2007SOR}}. The Jastrow factor was optimized only with VMC gradients, and it was held fixed during optimization with FN gradients. 
The FN gradients were computed from a standard walker distribution using mixed estimators, which corresponds to Method A in Ref.~\citenum{2022MCF}.
The lattice discretized version of the FN-DMC calculations (LRDMC)~{\cite{2005CAS, 2020NAK2}} was used in this study. The single-shot LRDMC calculations were performed by the single-grid scheme~{\cite{2005CAS}} with lattice spaces $a$ = 0.30, 0.25, 0.20, and 0.10 Bohr, and the energies were extrapolated to $a \rightarrow 0$ using $f(a^2)=k_4 \cdot a^4 + k_2 \cdot a^2 + k_0$.
%, where $k_0$ is the extrapolated energy. 
The LRDMC calculations for computing those gradients were performed by the single-grid scheme~{\cite{2005CAS}} with lattice spaces $a$ = 0.20 Bohr. The Determinant Locality approximation (DLA)~{\cite{2019ZEN}} was employed for the LRDMC calculations~\footnote{The use of DLA in LRDMC is equivalent to the DLTM~{\cite{2019ZEN}} scheme in standard DMC.}. We notice that the LRDMC framework guarantees the variational principle even with the presence of non-local pseudopotentials, as proven in the Appendix.
% VESTA
The molecular structures are depicted using VESTA~{\cite{2011MOM}}.

%{\vspace{3mm}}
\section{Results and Discussion}
%\paragraph{\textit{Results and Discussion} $-$}
%\subsection{Validations of the proposed scheme}

\subsection{The FNAGPAS captures strong correlation}
\begin{figure}[htbp]
  \centering
  \includegraphics[width=.5\columnwidth]{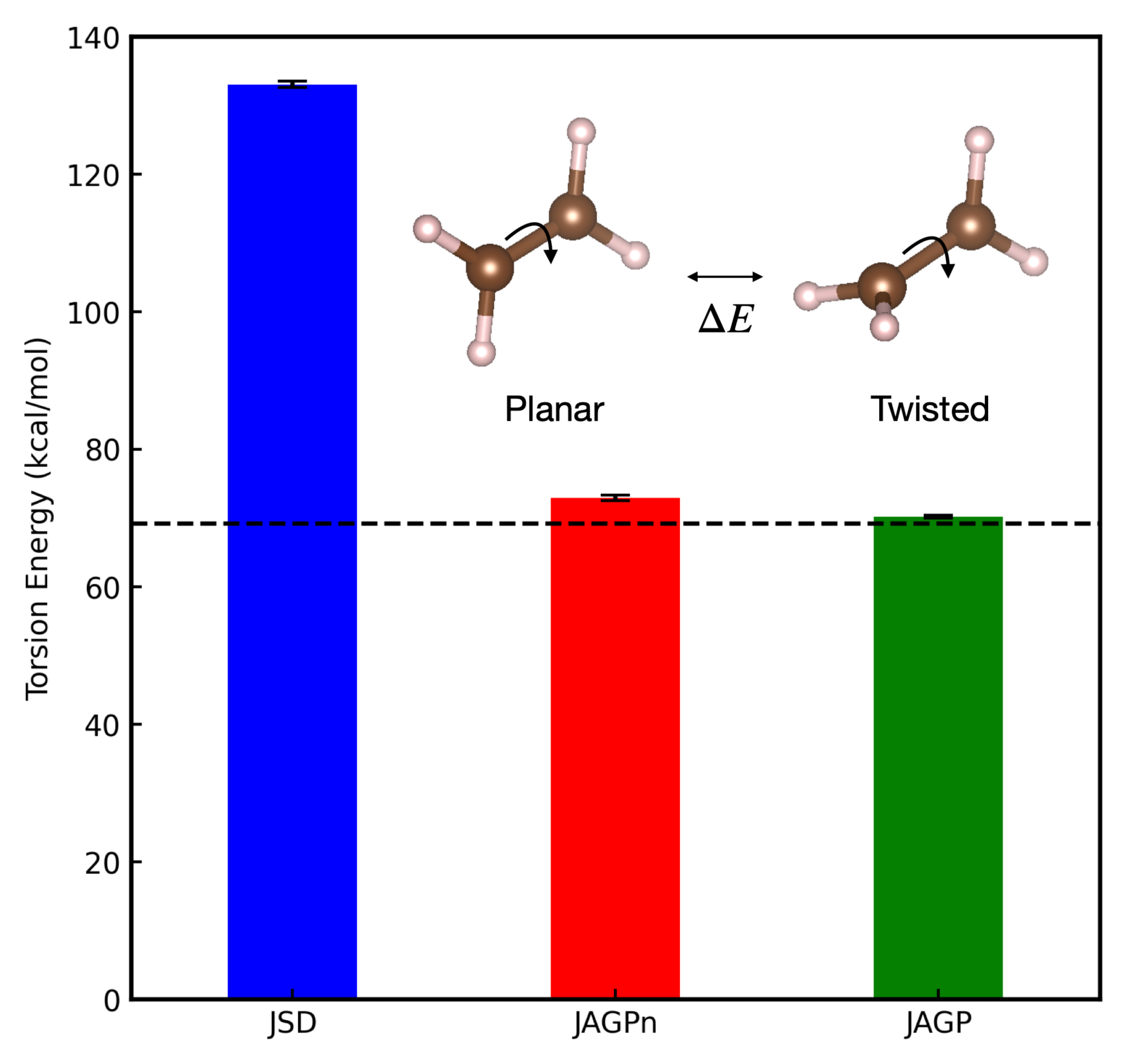}
  \caption{Torsion energies between the planar and twisted ethylene. The values of JAGP and MR-CISD+Q (horizontal broken line) are taken from Ref.~{\citenum{2014ZEN}} and Ref.~{\citenum{2004BAR}}, respectively.}
  \label{fig:ethylene-torsion}
\end{figure}

%\vspace{2mm}
% The nodes obtaind with AGPn ansatz opt. by LRDMC grads also imporive physical properties. Here, torsion energy of C2H4.
We show that the proposed FNAGPAS %AGPn ansatz obtained using FN gradients 
is able to incorporate the correlation effect that the JSD ansatz cannot do at all. 
We apply our scheme for the torsion energy estimation of ethylene (C$_2$H$_4$).
% def. torsion energy
The torsion energy is defined as the energy difference between the ground state ethylene structure (denoted as planar ethylene) and the orthogonally rotated ethylene structure (denoted as twisted ethylene), which are both shown in the inset of Fig.~\ref{fig:ethylene-torsion}.
Here, we consider only the singlet states for both configurations.
% JSD not good
It was shown~{\cite{2014ZEN}} that the JSD ansatz cannot describe the torsion energy correctly since the ansatz cannot consider the static electronic correlation of the twisted ethylene, which has a diradical character. 
This is true both at the variational and at the FN level of theory~{\cite{2014ZEN}}.
The lack of reliability in the FN results based on a JSD ansatz indicates that projection schemes cannot recover strong correlation if the FN constraints are given from a wave function with qualitative issues, due to the constraint on the projection coming from the trial wave function. Thus, the way to improve the quality of the FN results is to adopt a more general ansatz, able to improve the nodes of the trial wave function and enhance the reliability of FN estimations.

% why SD is not good?
The planar ethylene has an electronic structure characterized by a highest occupied molecular orbital (HOMO) of type $\pi$ and a lowest unoccupied molecular orbital (LUMO) of type $\pi^*$, and the HOMO-LUMO gap is finite. A single Slater determinant having two electrons of unlike spin on the HOMO and no electrons on the LUMO captures qualitatively well the nature of the wave function and there is no static correlation. 
However, when the molecule is twisted, the HOMO-LUMO gap decreases because the overlap between the $p$ orbitals (orthogonal to the plane of the $-\text{CH}_2$ atoms) of the two carbons decrease. At a torsional angle on 90 degrees (i.e., twisted ethylene) the two $p$ orbitals become orthogonal and the frontier orbitals become degenerate, forming two singly occupied molecular orbitals (SUMOs). 
We can define three independent (orthogonal) wave functions having two electrons on two degenerate orbitals forming a spin singlet, a diradical and two zwitterionic states \cite{Salem1972}.
Their wave functions imply the use of more than one Slater determinant, i.e., their electronic structure shows strong correlation.
Thus, a multireference ansatz is needed to correctly describe the diradical character of the orthogonally twisted ethylene~{\cite{2014ZEN}}.

%The twisted ethylene has two degenerated molecular orbitals, corresponding to the $\pi$ and $\pi^{*}$ of the planar  ethylene structure, which should be occupied with two electrons of same spin. This is not an electronic structure that can be defined with a single Slater determinant. A multireference ansatz, such as JAGP, is needed to correctly describe the diradical character of the orthogonally twisted ethylene~{\cite{2014ZEN}}.

% results
Figure~{\ref{fig:ethylene-torsion}} shows the torsion energies of ethylene computed with the JSD ansatz with a HF nodal surface, and the same energies computed with the JAGPn ansatz with HF molecular orbitals \footnote{ The HF orbitals obtained with the Fermi-Dirac smearing method were used for the occupied and the virtual orbitals of the JAGPn ansatz for the twisted ethylene, because the HOMO and LUMO should have the same energies. Note, in this case we did not use the natural orbitals (introduced in the discussion above), because this system is characterized by strong correlation coming from the two frontier orbitals, which are easily derived already from the HF theory. Moreover, in the twisted ethylene case we allowed the optimization of the off-diagonal coefficient of the AGP matrix that pairs the two frontier orbitals. }, whose weights are optimized using DMC gradients. As a comparison, we also show results obtained with the full JAGP ansatz optimized using VMC gradients, which was taken from Ref.~{\citenum{2014ZEN}}.  The reference value in Fig.~{\ref{fig:ethylene-torsion}} is taken from Ref.~{\citenum{2004BAR}}, and it is computed using MR-CISD+Q\footnote{ The twisted ethylene is a prototypical example of a system characterized by strong correlation where single reference perturbative approaches, such as CCSD(T), fail and multireferene approaches are needed. }. The JSD ansatz gives 133.1(4) kcal/mol for the torsion energy, which is far from the reference value obtained by MR-CISD+Q (i.e., 69.2 kcal/mol~{\cite{2004BAR}}). Our JAGPn ansatz gives a FN energy of 73.0(4) kcal/mol for the torsion energy, which is close to the reference values.
This result demonstrates that the JAGPn ansatz optimized using FN gradients correctly describes the diradical character of the orthogonally twisted ethylene, something that the JSD ansatz cannot do.

%%%%%%%%%%%%%%%%%%%%%%%%%%%%%%%%%%%%%%%%%%%%%%%%%%%%
%%%%%%%%%%%%%%%%%%%%%%%%%%%%%%%%%%%%%%%%%%%%%%%%%%%%

\subsection{Application of FNAGPAS to small and large systems}

\begin{figure}[htbp]
\centering
\includegraphics[width=.9\columnwidth]{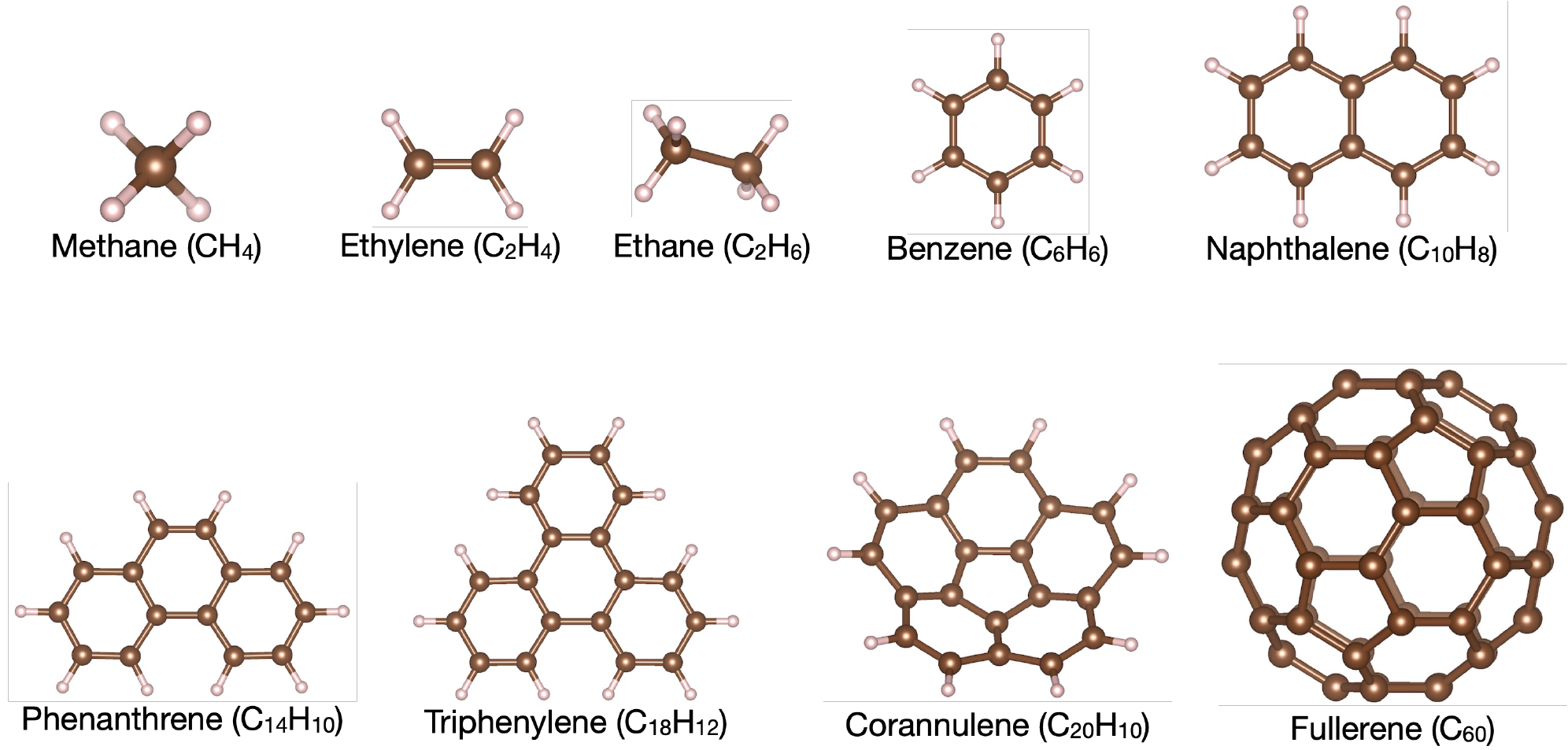}
\caption{
Molecular systems considered in this work, whose FN energy has been computed with the traditional JSD ansatz and with the JAGPm ansatz (within the FNAGPAS scheme) discussed in this work. 
The energy gain (i.e., the improvement of the FNAGPAS scheme over the traditional scheme which employs the JSD ansatz) and the number of variational parameters in the wave function for each system are shown in figure~\ref{fig:energy-gain}.
}
\label{fig:systems}
\end{figure}

We now show that the FNAGPAS scheme leads to a systematic improvement over the traditional JSD ansatz in molecular systems of increasing size, showing an accuracy in line with the full JAGP ansatz (and better on systems where the optimization error for the JAGP ansatz is large), while being affordable on much larger systems.
We consider the eight hydrocarbons and the C$_{60}$ fullerene, represented in Figure~\ref{fig:systems}.

%%%%%%%%%%%%%%%%%%%%%%%%%%%%%%%%%%%%%%%%%%%%%%%%%%%%
%%%%%%%%%%%%%%%%%%%%%%%%%%%%%%%%%%%%%%%%%%%%%%%%%%%%

\begin{figure}[htbp]
\centering
\includegraphics[width=.5\columnwidth]{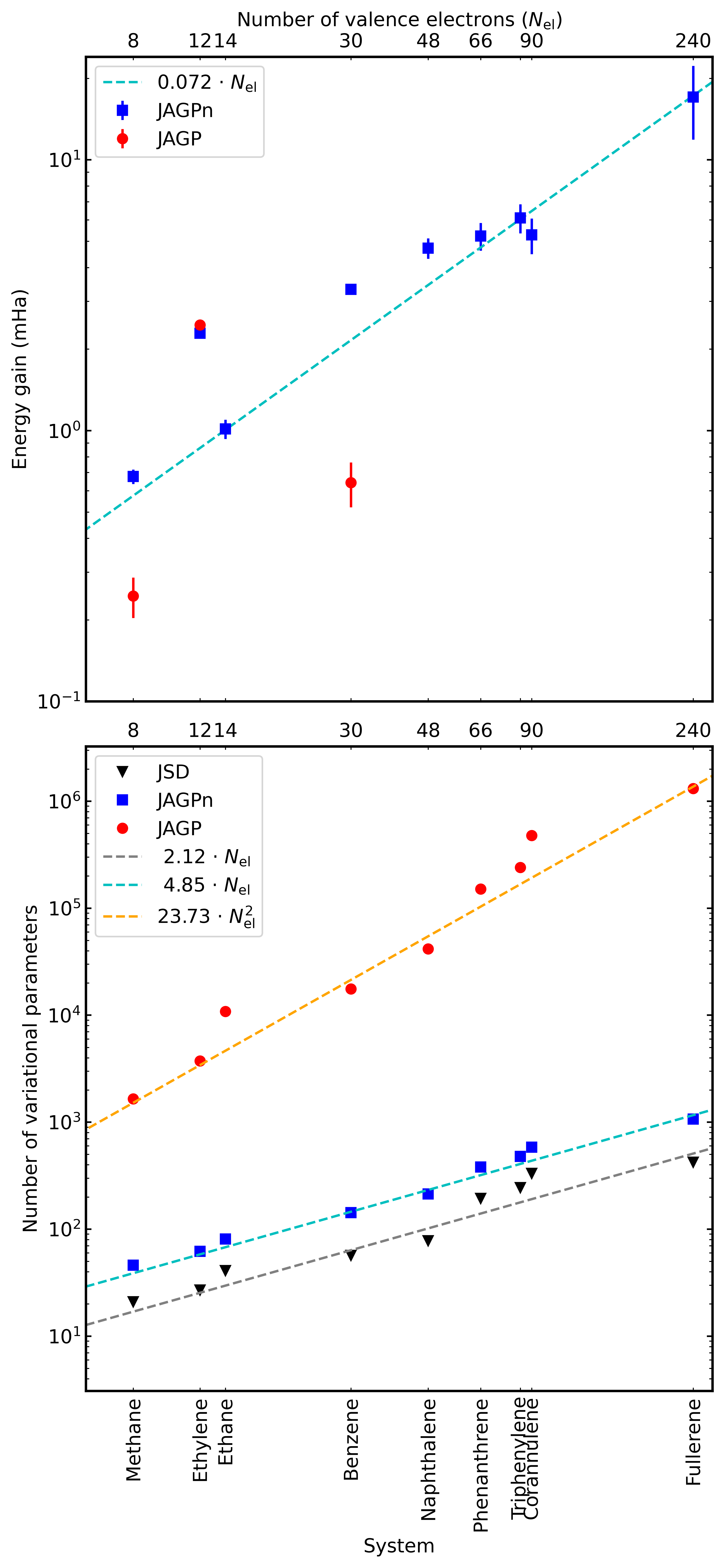}
\caption{
The top panel shows the improvement, dubbed energy gain, of the JAGP (red) and JAGPn (blue) ansatz with respect to the traditional JSD ansatz for each of the considered systems, as a function of the number of valence electrons. The energy gain is difference between FN energy of the JSD ansatz and the JAGP (or JAGPn) ansatz. 
The bottom panel shows the number of parameters in the Jastrow, in the AGP and in the AGPn parts of the wave function.  The dashed lines show the linear (gray for JSD, cyan for JAGPn) and quadratic (orange for JAGP) fitting curves.
}
\label{fig:energy-gain}
\end{figure}

% The nodes obtaind with AGPn ansatz opt. by LRDMC grads are bettenr than others, in terms of total energies.
Figure~{\ref{fig:energy-gain}} (top panel) shows the energy gain in the LRDMC total energies ($a \rightarrow 0$) by the nodal-surface optimizations of JAGP and JAGPn over the traditional JSD ansatz (with the nodal surface taken from the DFT LDA calculations). 
Our proposed FNAGPAS scheme (JAGPn ansatz optimized using FN gradients) shows positive gains for all molecules, indicating that the nodal surface optimizations improve the nodes of the Slater determinant obtained from DFT. Therefore, there is a systematic improvement in the description of the correlation energy. The energy gain scales linearly with the number of electrons in the system. 
% JAGP - JAGPn
The traditional JAGP ansatz (optimized using VMC gradients) 
%because the FN optimization was not stable) 
was computationally affordable only on the four smallest systems, due to the rapid increase of the number of variational parameters (see the bottom panel in Figure~\ref{fig:energy-gain}), which makes the optimization unstable or not converging.
In addition, we could only use VMC-opt for the JAGP ansatz, because FN-opt is not stable. This highlights an additional crucial advantage of FNAGPAS over the traditional JAGP approach.
In the four systems where we have both the traditional JAGP and the FNAGPAS results, the latter is equivalent to the former on ethane, and it recovers more correlation energy in methane, ethylene and benzene.
Larger systems were computationally unaffordable with JAGP, while JAGPn optimization remains feasible both at the variational and at the FN level. 
In fact, %a variational optimization of the nodal surface with the AGPn ansatz 
FNAGPAS has been successfully performed up to C$_{60}$ fullerene. The gain in C$_{60}$ is $\sim$ 2 meV/valence electron, as shown in the inset of Fig.~{\ref{fig:energy-gain}}. This is a reasonable value, considering a previous study by Marchi et al. reporting $\sim$ 3meV/valence electron for the finite-size graphene calculations with the same atoms as the C$_{60}$~{\cite{2011MAR}}.

%%%%%%%%%%%%%%%%%%%%%%%%%%%%%%%%%%%%%%%%%%%%%%%%%%%%
%%%%%%%%%%%%%%%%%%%%%%%%%%%%%%%%%%%%%%%%%%%%%%%%%%%%

% JAGPn has better gains
Let us consider more closely the medium-size molecules. Figure~\ref{fig:energy-gain} shows that the gains of JAGPn (optimized with FN gradients) are larger than those of JAGP (optimized using VMC gradients) in spite of the compactness of the AGPn ansatz. In fact, the number of variational parameters in the benzene molecule is 86 for the JAGPn ansatz, and is 17,629 for the JAGP ansatz. Moreover, JAGP is a generalization of JAGPn. Therefore, one could naively expect that the larger the number of variational parameters, the lower the energy. Here, we observe an exception to this expectation. For this point, we recall that the calculations reported in figure~\ref{fig:energy-gain} are obtained with a quite small Jastrow factor, employing a [3s1p] basis set for C atoms and a [3s] for H atoms. This is because we target large systems with FNAGPAS, for which the use of large Jastrow factors is unaffordable. It has been reported that an incomplete Jastrow factor leads to misdirection of the nodal surface within the variational optimization of the JAGP ansatz in the square H$_4$~{\cite{2019GEN}}. To confirm if this is the case in the present calculations, we performed additional calculations with a larger Jastrow factor in the JAGP ansatz calculations (i.e., a basis set of [4$s$3$p$1$d$] and [3$s$1$p$] for C and H atoms, respectively) and obtained that the larger Jastrow factor leads to a much larger energy gain than that obtained with the JAGP ansatz with a small Jastrow (see results in the SI (Table~S-I and Fig.~S-I). The result indicates that the small Jastrow factor leads to misdirection of the nodal surface of the JAGP ansatz also in this study.
% Which is the key? ansatz choice or gradient choice?
On the other hand, figure~\ref{fig:energy-gain} demonstrates that the FNAGPAS scheme works even with a small Jastrow factor and a minimal number of parameters in the antisymmetric part, making the approach applicable to larger systems. 

As mentioned in the method part,
see Section~\ref{sec:methods},
%{\bf Dario: is this a section? Can we refer to it?}, 
the two main features over which FNAGPAS is built are: 1) the AGPn ansatz, and 2) the optimization of its nodal surface using FN gradients. To reveal which of the two is more crucial for the success of the method, i.e., the ansatz or the gradient, we tried the following combinations: 
(i) JAGPn with VMC-opt; 
(ii) JAGPn with FN-opt, 
(iii) JAGP with VMC-opt;
(iv) JAGP with FN-opt.
Note that (ii) corresponds to FNAGPAS.
The scheme (iv), unfortunately, is not possible as the JAGP has too many parameters and the FN optimization becomes unstable.
Results obtained with schemes (i) to (iii) are reported in the SI (Table~S-I and Fig.~S-I).
We observe that scheme (ii) gives the best gains. Scheme (i) gives gains close to (ii), and they both are much better than (iii).
Thus, it emerges that freezing the orbitals to those obtained by a mean-field approach plays a crucial role in avoiding misdirection of the nodes optimization.

\subsection{The FNAGPAS scheme is size-consistent}

%\subsection{Binding energy of the methane--water dimer}
% The nodes obtaind with AGPn ansatz opt. by LRDMC grads also imporive physical properties. Here, binding energy (and the size-consistency).
%%%%%%%%%%%%%%%%%%%%%%%%%%%%%%%%%%%%%%%%%%%%%%%%%%%%
% Jastrow effect
\vspace{2mm}
We have shown that the AGPn ansatz is able to gain correlation energies at the FN level using very few variational parameters. 
In addition to their role in improving the nodal surface, FN gradients also appear to be crucial when calculating binding energies of molecules, preserving size consistency. As shown in Table~{\ref{tab:water-mathane-energy}} and discussed hereafter for the particular case of the water-methane dimer, this is not the case when VMC gradients are used.
Therefore, when calculating binding energies of molecules, the use of VMC gradients in the JAGPn ansatz gives incorrect results, while the use of FN gradients plays a crucial role on it.

%%%%%%%%%%%%%%%%%%%%%%%%%%%%%%%%%%%%%%%%%%%%%%%%%%%%
% Comparison of Water-Methane energies
%%%%%%%%%%%%%%%%%%%%%%%%%%%%%%%%%%%%%%%%%%%%%%%%%%%%
\begin{center}
\begin{table}[htbp]
\caption{\label{tab:water-mathane-energy}
FN binding energy $E_{\rm{b}}$ and size consistency energy error $E_{\text{SCE}}$, computed with LRDMC $a \rightarrow 0$, as obtained with the JSD, JAGPn and JAGP wave functions.
For JAGPn we consider both the case of using VMC and FN gradients to optimize the nodal surface. The latter is the scheme dubbed FNAGPAS in this work.
}
\vspace{2mm}
\begin{tabular}{c|c|c|c}
\Hline 
Ansatz & Nodes Opt. & $E_{\rm{b}}$ (meV) & $E_{\text{SCE}}$ (meV) \\ 
\Hline
   JSD &      - &                -27(2) &                                    -1(1) \\
 JAGPn & VMCopt &                -46(2) &                                    10(2) \\
 JAGPn &  FNopt &                -29(2) &                                    -2(2) \\
  JAGP & VMCopt &                -41(3) &                                    11(3) \\
\Hline
CCSD(T) & - &                    -27 &                                           0 \\
\Hline
\end{tabular}
\end{table}
\end{center}
%%%%%%%%%%%%%%%%%%%%%%%%%%%%%%%%%%%%%%%%%%%%%%%%%%%%

Table~{\ref{tab:water-mathane-energy}} contains the binding energies of the methane--water dimer computed with the JSD ansatz, with the JAGPn ansatz optimized using either VMC or FN gradients (the FNAGPAS approach), and with the JAGP ansatz optimized with VMC gradients. The binding energy is evaluated as the energy difference between the dimer and the sum of the energies of the two molecules: 
$E_{\rm{b}} = E_\text{water-methane} - E_\text{water} - E_\text{methane}$. The reference value for the binding energy of the water-methane dimer, -27~meV, was computed by CCSD(T) implemented in {\textsc{Orca}}~{\cite{Orca, Orca_v4}} program~{\footnote{
In particular, we performed canonical CCSD(T) calculations with the automatic basis set extrapolation implemented in Orca (which assumes an exponential convergence for the Hartree-Fock energy, and a polynomial convergence for the correlation energy) using Dunning correlation-consistent core-polarized basis sets, cc-pCV$n$Z, with quadruple-zeta ($n=\text{Q}$) and quintuple-zeta ($n=5$) basis set. We performed both estimations with and without counterpoise correction, both yielding a binding energy of -27.2~meV.
}. 
We chose the CCSD(T) value as a reference because the bounded water-methane dimer is not a strongly-correlated system, thus CCSD(T) should describe the binding energy correctly.
In this system the JSD ansatz gives a binding energy of $-27(2)$~meV, which is in good agreement with the CCSD(T) values of -27.0 meV. Thus, a new DMC approach with nodal surface optimization should lower the value of the total energies but should not affect the energy differences.
The FNAGPAS scheme, which optimizes 
% {\bf Dario: sometimes we say optimize and some others we say optimize. We should be consistent with either the British or with the American spelling} 
the JAGPn parameters with the FN gradients, behaves as expected, yielding a binding energy of -29(2)~meV, still in good agreement with the reference value.
However, this is not the case for the JAGPn ansatz optimized with the VMC gradients, which gives $E_{\rm{b}} = -46(2)$~meV, or for the JAGP ansatz (with VMC optimization), which gives $E_{\rm{b}} = -41(3)$~meV.

{\vspace{2mm}}
We can interpret the deterioration of the binding energy as follows: Binding energies are computed from relative energies among two or more molecules; thus, the accuracy relies on its error cancellation. The error cancellation in DMC was reviewed and discussed by Dubeck{\'y} in 2016~{\cite{Dubecky2016}}. Their conclusion is that one can rely on error cancellation as long as one keeps the constructions and optimizations of the corresponding wave functions as systematic as possible. Indeed, this cancellation works when the nodes are kept at the same systematic accuracy at every step of the trial wave function constructions. In fact, for the water-methane dimer calculations in this study, our JSD ansatz fully satisfies the size consistency and gives satisfactory binding energy, which means that the error cancellation works with the DFT nodal surfaces.
In this study, we found that error cancellation was deteriorated by the nodal surface optimizations using the VMC gradients while recovered by those using the FN-DMC gradients. When one computes the binding energy of a complex system, one usually uses the same Jastrow basis sets for each element in the complex and the isolated systems. The use of the same Jastrow basis sets does not guarantee the same contribution to the total energy of both the complex and the isolated systems at the VMC level. Indeed, during the nodal surface optimization at the VMC level, the incomplete Jastrow factor affects the nodal surface {\it differently} between the complex and isolated systems; thus, the resultant nodal surface gives the incorrect binding energy. The recovery should be because FN-DMC is a projection method to relax the amplitude of the AGPn ansatz, which corresponds to adding an unlimited flexible Jastrow factor to a given ansatz.

{\vspace{2mm}}
The Jastrow incompleteness is also related to the deterioration of the size consistency for JAGPn and JAGP with VMC optimization. The \emph{size consistency} is a property that guarantees the consistency of the energy behavior when the interaction between the involved molecular system is nullified (e.g., by a long distance). If the size consistency is fulfilled, the energy of the far-away system should be equal to the sum of the energies of the two isolated molecules. 
The last column in Table~{\ref{tab:water-mathane-energy}} shows the difference in energies of the faraway water-methane complex (at a distance of $\sim 11$~\AA) and the sum of the isolated molecules, which can be considered the size-consistency error and is here dubbed $E_{\text{SCE}}$. 
JSD ansatz is size consistent, as expected \cite{Zen2016}.
The table clearly shows that the size consistency is deteriorated by the optimization using VMC gradients, i.e., the difference between the isolated and far-away energies is finite. In contrast, the size consistency is perfectly retrieved by the optimization using FN gradients. 
Neuscamman~{\cite{Neuscamman2012}} pointed out that the deterioration of the size consistency comes from an incomplete Jastrow factor. More specifically, the real-space three/four-body Jastrow factor, which was employed in the present study, cannot completely remove the size consistency error unless we use unlimited flexibility in the Jastrow~{\cite{Neuscamman2012}}. To solve the problem, Goetz and Neuscamman proposed the so-called number-counting Jastrow factors that can suppress the unfavorable ionic terms, and is able to solve the size-consistency problem~{\cite{2017GOE, 2019GOE}} within the VMC framework.
In this regard, our proposed scheme can be interpreted as an alternative approach because, again, FN-DMC is a projection method to relax the {\it amplitude} of the AGPn ansatz, which corresponds to adding an unlimited flexible Jastrow factor to a given ansatz.

\subsection{Discussion}
First, we compare our approach with others that also target to go beyond the single-reference fixed-node approximation. A well-established strategy is to use the multi-determinant ansatz, which has witnessed numerous successes so far~{\cite{2010BOO, 2012PET, 2012MOR, 2015GIN, 2016MIC, 2020YAO, 2020SCE, 2020BEN, 2020MAL}}. 
The multi-determinant approach offers the advantage of systematic improvement by increasing the number of SDs. Nonetheless, the number of SDs for a comprehensive representation exponentially scales with system size, imposing substantial computational demands for large systems. 
Therefore, this method has mainly been applied to small molecular systems~\cite{2010BOO, 2012PET, 2012MOR}. 
However, there have been successful efforts to reduce the number of required determinants by neglecting less important ones~{\cite{2015GIN, 2016MIC}} using, for instance, the configuration interaction using a perturbative selection made iteratively (CIPSI), which mitigates the exponential character of the multi-determinant approach in practice~{\cite{2020SCE, 2020MAL}}. Recently, Benali et al. successfully applied the multi-determinant approach for solids with more than a hundred electrons by combining the CIPSI technique with a restricted active space built using natural orbitals~{\cite{2020BEN}}, which is a similar idea as we present in this study. Indeed, they demonstrated that one can go beyond the single-reference nodal surface in large systems by the multi-determinant approach in practice, though its naive asymptotic scaling is exponential. The multi-determinant approach is becoming as practical and promising as the single-determinant approach.

Concerning the actual computational costs of our proposed methods, the choice of ansatz (i.e., JSD or AGPn) does not significantly affect the cost of WF optimization, while the choice of gradients does. For instance, for C$_{60}$, Jastrow optimization with the JSD ansatz and Jastrow+nodal surface (i.e., weights of Natural Orbitals) optimization with the JAGPn ansatz using VMC gradients require 11.9 and 43.6 cores $\cdot$ hours per optimization step with $\sim$ 7 mHa accuracy on the total energy evaluation at each optimization step, respectively~{\footnote{We measured the computational times on the Numerical Materials Simulator at National Institute for Materials Science (NIMS) using 1536 cores (32 nodes $\times$ Intel Xeon Platinum 8268 (2.9GHz, 24cores) $\times$ 2 per node).}}. However, if one uses FN gradients for WF optimization, one needs more computational time. For instance, for C$_{60}$, the nodal surface (i.e., weights of Natural Orbitals) optimization with the JAGPn ansatz using FN gradients with $a$ = 0.20 a.u. requires 195.3 cores $\cdot$ hours per optimization step with $\sim$ 7 mHa accuracy on the total energy evaluation at each optimization step. Thus, our FNAGPAS scheme using FN gradients shows the same scaling of the number of variational parameters as the single-reference FN DMC with JSD ansatz while it increases the prefactor of computational cost.

%\vspace{2mm}
%
Based on the results obtained in this work so far, we finally discuss how to improve a fermionic ansatz in ab initio QMC calculations, in general. Recently, there have been many successful reports about machine-learning-inspired ansatz with a huge degree of freedom in describing electronic and spin states, such as deep neural networks~{\cite{2018CAR}}, restricted Boltzmann machines~{\cite{2017CAR, 2017NOM, 2021NOM}} and transformers~{\cite{2023VIT}}, which are utilized as ansatz of wave functions to solve the Schrödinger equation with lattice Hamiltonians. Also, in the ab initio community, ansatz using deep neural networks have been successfully applied for realistic problems, such as PauliNet~{\cite{PauliNet_Hermann2020}}, FermiNet~{\cite{FermiNet_Pfau2020}}, and others~\cite{Li2022, Ren2023, Choo2020, ForwardLaplacian_Li2023, Psiformer_VonGlehn2022}} . In light of the present results, let us consider exploiting an ansatz with a huge degree of freedom (i.e., many variational parameters) in ab initio QMC calculations to pursue an exact fermionic ground state. If we stop at the VMC level, we may apply such a flexible ansatz to Jastrow factors, determinant part, or both parts, and it is expected that the larger the degree of freedom an ansatz has, the larger the energy gain should be.
%, as seen in the JAGP calculations in this study. 
%However, if we proceed to FN-DMC, exploiting the flexibility for the Jastrow part seems to make more sense rather than for the determinant part. This is because, as revealed in this work, a larger degree of freedom in the determinant part does not necessarily guarantee a better nodal surface without a correct Jastrow factor. Moreover, if unlimitedly flexible Jastrow factors can be realized with a flexible ansatz, the VMC optimization is expected to fulfill the size consistency, which should be helpful for computing physical properties. \tobe{\bf DA I think the above paragraph could be confusing, as it is discussing the different point related to the presence of the Jastrow. }
However, improvements at the VMC level do not necessarily lead to improvements at the FN level, especially if the determinant part is optimized at the variational level.
A variational optimization improves the overall shape of the trial wave function $\Psi_{\text{T}}$, whilst the nodal surface might not be as optimized as the $\Psi_{\text{T}}$.
In this work, indeed, we have seen how the JAGPn ansatz optimized at the FN level leads to much better results than the JAGP ansatz optimized at the VMC level, despite the latter having many more variational parameters and it is much better at the VMC level. Moreover, we also observed how the JAGPn (and JAGP, for that matter) ansatz itself yields a size-consistency error at the FN level if the parameters are optimized at the VMC level, while the same ansatz with parameters optimized at the FN level is not affected by this issue.
Thus, caution should be used when employing these new highly flexible machine-learning-based wave function parametrizations, as it is not guaranteed that improvements in the VMC energy are reflected in improvements in the FN energy in a consistent way. Basic physical properties, which were present in the most standard wave functions (such as the JSD), might not appear in the more fancy approaches, similar to the mentioned problem of size-inconsistency in JAGP and JAGPn.

\section{Conclusions and perspectives}
\vspace{2mm}
In this study, we propose a method for variational optimization of the AGP wavefunction expressed in terms of natural orbitals, with pairing coefficients optimized using FN gradients. Within our scheme, the variational parameter space increases only linearly with the system size, as opposed to the quadratic scaling of the standard parametrization of AGP, with the result that our proposed method allows the optimization of the nodal surfaces for large systems, which has been difficult to achieve with conventional approaches. In addition to demonstrating that our scheme can be applied to systems as large as C$_{60}$, we showed that our scheme also achieves better (i.e., lower) DMC energies than the single-reference fixed-node DMC calculations. Moreover, we have shown that our approach is size-consistent and can be used to estimate binding energies.

We showed that the Jastrow incompleteness affecting nodal surface optimizations can be mitigated by using FN gradients combined with the JAGPn ansatz. However, in this study, we did not investigate the effect of the basis set incompleteness on the determinant part (i.e., nodal surface). The basis set incompleteness is believed to be less severe in QMC calculations than in quantum chemistry methods because the Jastrow factor (at the variational level) or the projection (at the FN level) mitigates its error. However, to the best of our knowledge, no one has seriously investigated the error so far. Considering binding energy calculations done by DMC reported so far~{\cite{Dubecky2016}, the basis set incompleteness should have a small effect on small molecules, but it should be carefully considered when studying large molecules using DMC done with localized basis sets. This is one of the intriguing future works for applying the single-reference-DMC and our proposed methods to large systems.

%
%Our work provides a clue on how the QMC community designs a new ansatz with more degrees of freedom in the future when they attempt to use DMC methods to get closer to exact solutions for fermions.

\section*{Acknowledgments}
%\vspace{2mm}
%\paragraph{{\textit{Acknowledgements}} $-$}
% NIMS computer
K.N. is grateful for computational resources from the Numerical Materials Simulator at National Institute for Materials Science (NIMS).
% FUGAKU
K.N. is grateful for computational resources of the supercomputer Fugaku provided by RIKEN through the HPCI System Research Projects (Project IDs: hp220060 and hp230030).
% K.N. financial support
K.N. acknowledges financial support from the JSPS Overseas Research Fellowships, from Grant-in-Aid for Early Career Scientists (Grant No.~JP21K17752), from Grant-in-Aid for Scientific Research (Grant No.~JP21K03400), and from MEXT Leading Initiative for Excellent Young Researchers (Grant No.~JPMXS0320220025).
D.A. and A.Z. acknowledge support from Leverhulme grant no. RPG-2020-038.
%Finanziato dall'Unione Europea -- Next generation EU
% Progetto AZ P2022MC742 
% Progetto DA 20222FXZ33
D.A. and A.Z. also acknowledge support from the European Union under the Next generation EU (projects 20222FXZ33 and P2022MC742).
The authors acknowledge the use of the UCL Kathleen High Performance Computing Facility (Kathleen@UCL), and associated support services, in the completion of this work. 
This research used resources of the Oak Ridge Leadership Computing Facility at the Oak Ridge National Laboratory, which is supported by the Office of Science of the U.S. Department of Energy under Contract No. DE-AC05-00OR22725). Calculations were also performed using the Cambridge Service for Data Driven Discovery (CSD3) operated by the University of Cambridge Research Computing Service (www.csd3.cam.ac.uk), provided by Dell EMC and Intel using Tier-2 funding from the Engineering and Physical Sciences Research Council (capital grant EP/T022159/1 and EP/P020259/1). This work also used the ARCHER UK National Supercomputing Service (https://www.archer2.ac.uk), the United Kingdom Car Parrinello (UKCP) consortium (EP/F036884/1).

\vspace{2mm}
% For Sandro
We dedicate this paper to one of the authors, Prof. Sandro Sorella (SISSA), who passed away during the collaboration. He initially devised an idea to use FN gradients for WF optimizations, which is one of the keys for the success of the present work. 
The QMC community will remember that he is one of the most influential contributors of the past and of the beginning of the present century to the community, and in particular for deeply inspiring this work with his developing of the {\emph{ab initio}} QMC package, \tvb~{\cite{2020NAK}}.

%\section{Data availability} \label{sec:data}
%\vspace{2mm}
%\paragraph{{\textit{Data availability}} $-$}
%The \trexio\ files containing the trial WFs used in this study are available from the ZENODO repository~{\color{red}[XXX]}.

\appendix
\section*{Appendix: Proof for the variational principle of the LRDMC optimzation with DLA} 
%\label{sec:appendix}

% (2) optimization
%\subsection*{Proof for the variational principle of the LRDMC optimzation with DLA}
%\label{sec:energy-minimization}
%
As pointed out in seminal works by Casula et al.~{\cite{2005CAS, 2006CAS}}, the use of a pseudo potential that has the so-called non-local term induces an additional sign problem in the standard DMC approach with the locality approximation (LA); thus the variatioanl principle, which justifies the energy minimization strategy, is deteriorated. Instead, one of the advantages of the LRDMC is that the use of pseudo potentials does not deteriorate the variational princple~{\cite{2005CAS}}; thus, the energy-minimization is justified. Recently, we implemented the Determinant Locality approximation (DLA)~{\cite{2019ZEN}} into the \tvb\ package. In this study, we combine the DLA with the LRDMC framework implemented in the \tvb\ package.
We prove here that the variational principle holds also in the LRDMC with the DLA.
This proof is inspired by the proof by \citet{1995HAA} that the lattice Green’s function Monte Carlo method is variational.

In LRDMC calculations with the DLA, the effective Hamiltonian (i.e., the fixed-node Hamiltonian) reads:
\begin{equation}
  H^{\rm FN}_{x,x'}=
  \begin{cases}
    H_{x,x} + V^{\rm sf, DLA}_{x,x} & \text{for $x'=x$} \\
    0                 & \text{for $x' \ne x$, if $\Psi_{\text{T}}(x')H_{x,x'}\Psi_{\text{T}}(x) > 0$,} \\
    H_{x,x'}       & \text{for $x' \ne x$, else}
  \end{cases}
\end{equation}
where $H_{x',x} \equiv \braket{x'|\hat{H}|x}$, $V^{\rm sf, DLA}_{x,x} = \sum_{x' \ne x}^{\text{for sf}} D_{\text{T}}(x')H_{x,x'}/D_{\text{T}}(x)$, by which the original term in the LRDMC approach, $V^{\rm sf}_{x,x} = \sum_{x' \ne x}^{\text{for sf}} \Psi_{\text{T}}(x')H_{x,x'}/\Psi_{\text{T}}(x)$, is replaced, and sf means that all $x' (\ne x)$ satisfying $\Psi_{\text{T}}(x')H_{x,x'}\Psi_{\text{T}}(x) > 0$.
Here, we omit the lattice-space dependency of the Hamiltonian (i.e., $H \equiv H^{a}$) because one can extrapolate energies to the $a \rightarrow 0$ limit. Notice that, we assume that a trial WF can be decomposed into the Jastrow and determinant parts, i.e., $\Psi_{\text{T}} = J_{\text{T}} D_{\text{T}}$. 
We also notice that 
$$\Psi_{\text{T}}(x')H_{x,x'}\Psi_{\text{T}}(x) > 0  \leftrightarrow D_{\text{T}}(x')H_{x,x'}D_{\text{T}}(x) > 0$$ 
since the Jastrow factor does not affect the sign of a wavefunction. 
We define the following notations:
\begin{equation}
E_{\rm MA}= {\braket{\Psi_{\rm T}|\hat{H}^{\rm FN}|\Phi_{\rm FN}} \over \braket{\Psi_{\rm T}|\Phi_{\rm FN}} }
\end{equation}
\begin{equation}
E_{\rm FN}={ \braket{\Phi_{\rm FN}|\hat{H}^{\rm FN}|\Phi_{\rm FN}} \over \braket{\Phi_{\rm FN}|\Phi_{\rm FN}} }
\end{equation}
\begin{equation}
E={\braket{\Phi_{\rm FN}|\hat{H}|\Phi_{\rm FN}}\over\braket{\Phi_{\rm FN}|\Phi_{\rm FN}} }
\end{equation}
\begin{equation}
E_{0}={\braket{\Psi_{0}|\hat{H}|\Psi_{0}} \over \braket{\Psi_{0}|\Psi_{0}} }
\end{equation}
where $\ket{\Phi_{\rm FN}}$ is the fixed-node ground state of $\hat{H}^{\rm FN}$.
In the following, we will show the following equations hold:
\begin{equation}
E_{\rm MA} = E_{\rm FN} \ge E \ge E_{0}.  
\end{equation}
The first equal ($E_{\rm MA} = E_{\rm FN}$) holds because $\ket{\Phi_{\rm FN}}$ is the exact ground state of $H^{\rm FN}$ (i.e. $\hat{H}^{\rm FN}\ket{\Phi_{\rm FN}} = E_{\rm FN}\ket{\Phi_{\rm FN}}$). This is also true with the non-local terms of pseudo potentials.
%
%Since $\ket{\Psi_{\rm eff}}$ is the fixed-node ground state of $H$ and is independent of the Jastrow factor of the trial WF $\Psi_{\rm T}$, neither $E^{\rm eff}_{\rm MA}$ nor $E^{\rm eff}_{\rm FN}$ is dependent on the Jastrow factor.
%
Now we define the difference between the effective fixed-node energy obtained with the effective Hamiltonian $\hat{H}^{\rm FN}$ and that obtained with the true Hamiltonian $\hat{H}$:
\begin{equation}
\Delta E \equiv E_{\rm FN} - E = \braket{\Phi_{\rm FN}|\hat{H}^{\rm FN} - \hat{H}|\Phi_{\rm FN}}.
\end{equation}
We want to prove that $\Delta E \ge 0$ for the fixed-node state and the equal holds for $\Phi_{\rm FN} = \Psi_{\rm T} = \Psi_{0}$, where we denote $\Psi_{0}$ as the exact WF of the original Hamiltonian $\hat{H}$, i.e., $\hat{H}\ket{\Psi_{0}} = E_0\ket{\Psi_{0}}$.
Hereafter, we will do the same exercise written in Ref.~{\citenum{1995HAA}}.
We define the difference between the effective fixed-node energy obtained with the effective Hamiltonian $H^{\rm FN}$ and that obtained with the true Hamiltonian $H$:
\begin{equation}
\Delta E \equiv E_{\rm FN} - E = \braket{\Phi_{\rm FN}|H^{\rm FN} - H|\Phi_{\rm FN}} = \braket{\Phi_{\rm FN}|V^{\rm sf} - H^{\rm sf}|\Phi_{\rm FN}},
\end{equation}
where, we define a truncated Hamiltonian $H^{\rm tr}$ and a spin-flip Hamiltonian $H^{\rm sf}$, by
\begin{equation}
H = H^{\rm tr} + H^{\rm sf}    
\end{equation}
and
\begin{equation}
H^{\rm FN} = H^{\rm tr} + V^{\rm sf}.
\end{equation}
Indeed, the matrix elements are:
\begin{equation}
  H^{\rm tr}_{x,x'}=
  \begin{cases}
    H_{x,x}       & \text{for $x'=x$} \\
    0                 & \text{for $x' \ne x$, if $\Psi_{\text{T}}(x')H_{x,x'}\Psi_{\text{T}}(x) > 0$,} \\
    H_{x,x'}      & \text{for $x' \ne x$, else}
  \end{cases}
\end{equation}
and
\begin{equation}
  H^{\rm sf}_{x,x'}=
  \begin{cases}
    0                 & \text{for $x'=x$} \\
    H_{x,x’}      & \text{for $x' \ne x$, if $\Psi_{\text{T}}(x')H_{x,x'}\Psi_{\text{T}}(x) > 0$,} \\
    0       & \text{for $x' \ne x$, else}
  \end{cases}
\end{equation}
$\Delta E$ can be written explicitly in terms of the matrix elements:
\begin{equation}
\Delta E = \sum_{x} \Phi_{\rm FN}^{*}(x) \left[
\braket{x|V^{\rm sf}|x}\Phi_{\rm FN}(x) -
\sum_{x'} \braket{x|H^{\rm sf}|x'}\Phi_{\rm FN}(x')
\right] \,,
\end{equation}
and rewriten as:
\begin{equation}
\Delta E = \sum_{x} \Phi_{\rm FN}^{*}(x) \left[
\sum_{x'}^{\rm sf}{
H_{x,x'}\frac{D_{\text{T}}(x')}{D_{\text{T}}(x)}\Phi_{\rm FN}(x)
}-
\sum_{x'}^{\rm sf}{
H_{x,x'}\Phi_{\rm FN}(x')
}
\right] 
\end{equation}
where, sf means that all $x' (\ne x)$ satisfying $\Psi_{\text{T}}(x')H_{x,x'}\Psi_{\text{T}}(x) > 0$.
In this double summation, each pair of configurations $(x, x')$ appear twice. Therefore, we can combine these terms and rewrite it as a summation over the pairs:
\begin{equation}
\Delta E = \sum_{(x, x')}^{\rm sf} H_{x,x'} \left[
\frac{D_{\text{T}}(x')}{D_{\text{T}}(x)}|\Phi_{\rm FN}(x)|^2
+
\frac{D_{\text{T}}(x)}{D_{\text{T}}(x')}|\Phi_{\rm FN}(x')|^2
-
\Phi_{\rm FN}^*(x)\Phi_{\rm FN}(x')
-
\Phi_{\rm FN}^*(x')\Phi_{\rm FN}(x)
\right] 
\end{equation}
Notice that the hamiltonian is hermitian: $H_{x,x'} = H_{x',x}$. 
Since all the pair satisfies $\cfrac{D_{\text{T}}(x')}{D_{\text{T}}(x)} H_{x,x'} > 0 \,,$ then, 
\begin{equation}
H_{x,x'} \frac{D_{\text{T}}(x')}{D_{\text{T}}(x)}
=
|H_{x,x'}| \left|\frac{D_{\text{T}}(x')}{D_{\text{T}}(x)}\right| \,\, {\text {and}} \,\,
H_{x,x'} \frac{D_{\text{T}}(x)}{D_{\text{T}}(x')}
=
|H_{x,x'}| \left|\frac{D_{\text{T}}(x)}{D_{\text{T}}(x')}\right|
\end{equation}
Then,
%
%\begin{equation}
\begin{multline}
\Delta E = \sum_{(x, x')}^{\rm sf} |H_{x,x'}| 
\biggr [
\left|\frac{D_{\text{T}}(x')}{D_{\text{T}}(x)}\right||\Phi_{\rm FN}(x)|^2
+ \left|\frac{D_{\text{T}}(x)}{D_{\text{T}}(x')}\right||\Phi_{\rm FN}(x')|^2
\\
- sgn(x, x') \Phi_{\rm FN}^*(x)\Phi_{\rm FN}(x')
- sgn(x, x') \Phi_{\rm FN}^*(x')\Phi_{\rm FN}(x)
\biggr ] 
\end{multline}
%\end{equation}
%
where $sgn(x, x')$ denotes the sign of $H_{x,x'}$. Finally, we get:
\begin{equation}
\Delta E = \sum_{x, x'}^{\rm sf} |H_{x, x'}| \left|\Phi_{\rm FN}(x)\sqrt{|\frac{D_{\text{T}}(x')}{D_{\text{T}}(x)}|} - sgn(x,x')\Phi_{\rm FN}(x')\sqrt{|\frac{D_{\text{T}}(x)}{D_{\text{T}}(x')}|}\right |^2,
\end{equation}
indicating that $\Delta E$ is positive for any wave function $\Phi_{\rm FN}$. Thus, the ground-state energy of $H^{\rm FN}$ is an upper bound for the ground-state energy of the original Hamiltonian $H$ (i.e., $E_{\rm FN} \ge E$).
Hereafter, we consider the case that one uses the true ground-state $\Psi_{0}$ for the determinant of the trial wave-function (i.e., $\Psi_{\text{T}} = J_{\text{T}} \cdot \Psi_{0}$), to prove that $E_{\rm FN} = E$ holds with $\Psi_{\text{T}} = J_{\text{T}} \cdot \Psi_{0}$ (i.e., $D_{\text{T}} = \Psi_{0}$):
For all the pairs $(x, x')$, $\Psi_{0}H_{x,x'}\Psi_{0} > 0$ is satisfied, meaning $sgn(x,x') \rightarrow + $ and $ \cfrac{\Psi_{0}(x)}{\Psi_{0}(x')} \rightarrow + $, or $sgn(x,x') \rightarrow - $ and $ \cfrac{\Psi_{0}(x)}{\Psi_{0}(x')} \rightarrow - $.
Thus, the above condition is fulfilled when the following condition is satisfied:
\begin{equation}
\frac{\Phi_{\rm FN}(x)}{\Phi_{\rm FN}(x')} = \frac{\Psi_{0}(x)}{\Psi_{0}(x')}.
\end{equation}
In the DLA approach, the spin-flip term is composed only of the determinant of the trial WF. Therefore, the fixed-node outcome with the DLA approach is not affected by the presence of the Jastrow factor in the trial WF (in the $a \rightarrow 0$ limit). 
Therefore, one gets $\Phi_{\rm FN} = \Psi_{0}$ with $\Psi_{\text{T}} = J_{\text{T}} \cdot \Psi_{0}$. Thus, $\Delta E = 0$ is fulfilled with $\Psi_{\text{T}} = J_{\text{T}} \cdot \Psi_{0}$, and the following relations hold:
\begin{equation}
E_{\rm MA} = E_{\rm FN} = E = E_{0}~(\text{with $\Psi_{\text{T}} \equiv J_{\text{T}} \cdot \Psi_{0}$}),
\end{equation}
meaning that the effective Hamiltonian $\hat{H}^{\rm FN}$ and the true Hamiltonian $\hat{H}$ has the same ground-state energy $E_0$ and the same ground state $\Phi_{\rm FN} = \Psi_{0}$ with $\Psi_{\text{T}} = J_{\text{T}} \cdot \Psi_{0}$, where the final equal $E = E_{0}$ comes from the usual variational principle. 

{\vspace{2mm}}
In the DLA approach, we can update the trial WF $\Psi_{\text{T}}$ such that $E_{\rm MA}$ goes down according to the gradient $\partial_\alpha E_{\rm MA}$ or using a more sophisticated optimization scheme. As written above, the equals $E_{\rm FN} = E = E_0$ are met when $\Psi_{\text{T}} = J \cdot \Psi_{0}$. It implies that one can look for the true ground-state energy and wavefunction by variation of the {\it determinant part} of the trial wavefunction.  Indeed, in the LRDMC calculations with the DLA, one can access the mixed-average energy $E_{\rm MA}$, and its derivative $\partial_{\vec{\alpha}} E_{\rm MA}$, where ${\vec{\alpha}}$ is a set of the variational parameters. Since $E_{\rm MA}$ satisfies the variational principle, i.e., $E_{\rm MA} \ge E_{0}$, the equal holds when $\Psi_{\text{T}} = J_{\text{T}} \cdot \Psi_{0}$, as proven above, one can update the {\it determinant part} of the trial WF, $D_{\text{T}}$, such that $E_{\rm MA}$ goes down, then, it is expected that $D_{\text{T}}$ finally reaches $D_{\text{T}} \rightarrow \Psi_{0}$, and $E_{\rm MA} \rightarrow E_{0}$.

\section*{Code availability} \label{sec:code}
%\vspace{2mm}
%\paragraph{{\textit{Code availability}} $-$}
The QMC kernel used in this work, \tvb, is available from its GitHub repository [\url{https://github.com/sissaschool/turborvb}].

\section*{Supporting Information Available} \label{sec:code}
%\vspace{2mm}
%\paragraph{{\textit{Code availability}} $-$}
We provide the total energies and the number of variational parameters of of hydrocarbons and fullerene.
We also report total energies of the methane-water complex and corresponding fragments, and of the planar and twisted ethylene molecule. We discuss the role of molecular orbitals or natural orbitals in the AGPn ansatz and the size-consistency error. 
The geometries, in xyz format, of all the systems studied are also provided.
This information is available free of charge via the Internet at http://pubs.acs.org

%%%%%%%%%%%%%%%%%%%%%%%%%%%%%%%
%\bibliographystyle{apsrev4-1}
\bibliography{./references.bib}
%%%%%%%%%%%%%%%%%%%%%%%%%%%%%%%

\end{document}

\title{Supplementary Material for Beyond single-reference fixed-node approximation in {\emph {ab initio}} Diffusion Monte Carlo using antisymmetrized geminal power applied to systems with hundreds of electrons} 
\author{Kousuke Nakano}
\email{kousuke\_1123@icloud.com}
\affiliation{Center for Basic Research on Materials, National Institute for Materials Science (NIMS), Tsukuba, Ibaraki 305-0047, Japan}
\affiliation{International School for Advanced Studies (SISSA), Via Bonomea 265, 34136, Trieste, Italy}
\author{Sandro Sorella}
\affiliation{International School for Advanced Studies (SISSA), Via Bonomea 265, 34136, Trieste, Italy}
\author{Dario Alf{\`e}}
\affiliation{Dipartimento di Fisica Ettore Pancini, Universit\`a di Napoli Federico II, Monte S. Angelo, I-80126 Napoli, Italy}
\affiliation{Department of Earth Sciences, University College London, Gower Street, London WC1E 6BT, United Kingdom}
\affiliation{Thomas Young Centre and London Centre for Nanotechnology, 17-19 Gordon Street, London WC1H 0AH, United Kingdom}
\author{Andrea Zen}
\email{andrea.zen@unina.it}
\affiliation{Dipartimento di Fisica Ettore Pancini, Universit\`a di Napoli Federico II, Monte S. Angelo, I-80126 Napoli, Italy} 
\affiliation{Department of Earth Sciences, University College London, Gower Street, London WC1E 6BT, United Kingdom}

\date{\today}

\begin{abstract}
This file includes the supplementary information for the paper titled ``Beyond single-reference fixed-node approximation in {\emph {ab initio}} Diffusion Monte Carlo using antisymmetrized geminal power applied to systems with hundreds of electrons".
\end{abstract}
\maketitle

\makeatletter
\def\Hline{
\noalign{\ifnum0=`}\fi\hrule \@height 1pt \futurelet
\reserved@a\@xhline}
\makeatother

\makeatletter
\renewcommand{\refname}{}
\renewcommand*{\citenumfont}[1]{#1}
\renewcommand*{\bibnumfmt}[1]{[#1]}
\makeatother

\setcounter{table}{0}
\setcounter{equation}{0}
\setcounter{figure}{0}
\renewcommand{\thepage}{S\arabic{page}}
\renewcommand{\thetable}{S-\Roman{table}}
\renewcommand{\thefigure}{S-\arabic{figure}}
\renewcommand{\theequation}{S-\arabic{equation}}
\renewcommand{\thelstlisting}{S-\arabic{lstlisting}}

\section{Supporting Results}

\subsection{Total energies and the number of variational parameters of hydrocarbons and fullerene}
\begin{figure}[htbp]
  \centering
  \includegraphics[width=0.7\columnwidth]{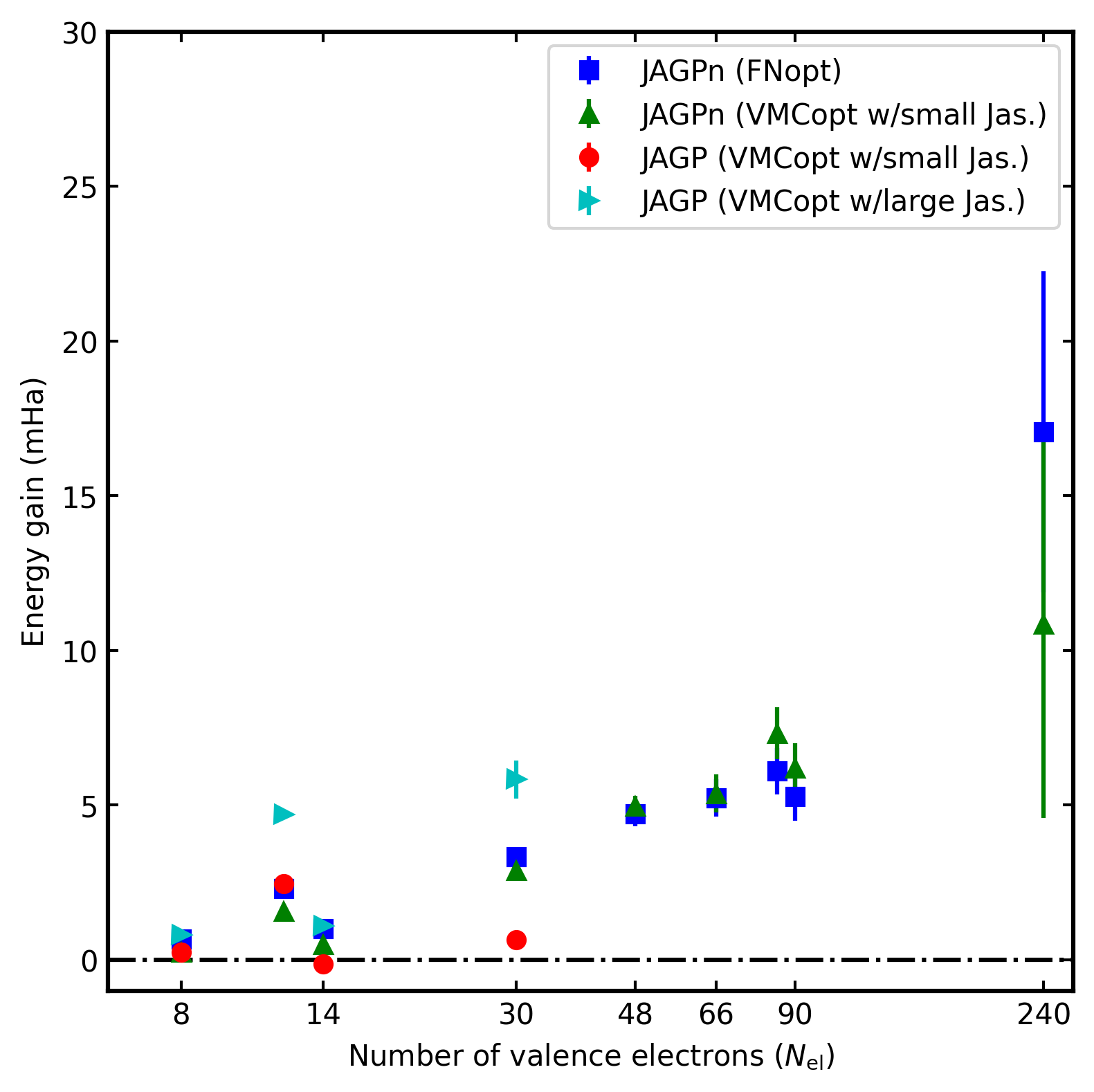}
  \caption{Improvements in the LRDMC energies ($a \rightarrow 0$), dubbed energy gain, of the JAGPn optimized with FN gradients (blue), JAGPn optimized with VMC gradients in the presence of the small Jastrow factor (green), JAGP optimized with VMC gradients in the presence of the small Jastrow factor (red), and JAGP optimized with VMC gradients in the presence of the large Jastrow factor (cyan) ansatze with respect to the traditional JSD ansatz for each of the considered systems, as a function of the number of valence electrons.}
  \label{fig:molecules-lrdmc-energy}
\end{figure}

Table~{\ref{tab:molecules-lrdmc-energy}} shows the LRDMC total energies of eight hydrocarbons (CH$_4$, C$_2$H$_4$, C$_2$H$_6$, C$_6$H$_6$, C$_{10}$H$_8$, C$_{14}$H$_{10}$, C$_{18}$H$_{12}$, C$_{20}$H$_{10}$), and the C$_{60}$ fullerene, obtained with different types of ansatz. Figure~{\ref{fig:molecules-lrdmc-energy}} plots the LRDMC energy gains with respect to that obtained with the corresponding JSD ansatz. Table~{\ref{tab:molecules-num-parameter}} shows the numbers of valence electrons and variational parameters for the 8 hydrocarbons and the C$_{60}$ fullerene.

%%%%%%%%%%%%%%%%%%%%%%%%%%%%%%%%%%%%%%%%%%%%%%%%%%%%
%\begin{sidewaystable}
\begin{table}[htbp]
\begin{center}
\caption{\label{tab:molecules-lrdmc-energy} The total energies obtained by LRDMC with DLA ($a \rightarrow 0$) calculations for the 8 hydrocarbons (CH$_4$, C$_2$H$_4$, C$_2$H$_6$, C$_6$H$_6$, C$_{10}$H$_8$, C$_{14}$H$_{10}$, C$_{18}$H$_{12}$, C$_{20}$H$_{10}$) and the C$_{60}$ fullerene. The units are in Hartree.}
\begin{tabular}{c|ccccccccc}
\Hline
Ansatz &       CH$_4$ &    C$_2$H$_4$ &    C$_2$H$_6$ &    C$_6$H$_6$ & C$_{10}$H$_8$ & C$_{14}$H$_{10}$ & C$_{18}$H$_{12}$ & C$_{20}$H$_{10}$ &     C$_{60}$ \\
\Hline
JSD, VMCopt~{\footnotemark[1]}                     &  -8.07878(3) &  -13.71162(1) &  -14.95337(6) &  -37.61952(4) &   -61.5103(3) &      -85.4062(4) &     -109.2998(4) &     -119.3914(4) &  -339.926(4) \\
JAGPn, FNopt~{\footnotemark[1]}                    &  -8.07945(3) &  -13.71391(1) &  -14.95438(6) &  -37.62284(5) &   -61.5150(3) &      -85.4115(4) &     -109.3059(6) &     -119.3967(7) &  -339.943(4) \\
JAGPn, VMCopt~{\footnotemark[1]}                    &  -8.07903(3) &  -13.71319(1) &  -14.95387(6) &   -37.6224(1) &   -61.5153(2) &      -85.4116(4) &     -109.3071(7) &     -119.3976(7) &  -339.937(5) \\
JAGP, VMCopt~{\footnotemark[1]}                    &  -8.07902(3) &  -13.71408(4) &  -14.95323(6) &   -37.6202(1) &             - &                - &                - &                - &            - \\
\Hline
JSD, VMCopt~{\footnotemark[2]}  &  -8.07869(7) &  -13.71165(8) &  -14.95331(5) &   -37.6199(4) &             - &                - &                - &                - &            - \\
JAGP, VMCopt~{\footnotemark[2]} &  -8.07949(7) &  -13.71634(8) &   -14.9544(2) &   -37.6257(4) &             - &                - &                - &                - &            - \\
\Hline
\end{tabular}
\footnotetext[1]{The small Jastrow basis sets ([3$s$1$p$] and [3$s$] for C and H atoms, respectively) are used.}
\footnotetext[2]{The large Jastrow basis sets ([4$s$3$p$1$d$] and [3$s$1$p$] for C and H atoms, respectively) are used.}
\end{center}
\end{table}
%\end{sidewaystable}
%%%%%%%%%%%%%%%%%%%%%%%%%%%%%%%%%%%%%%%%%%%%%%%%%%%%

%%%%%%%%%%%%%%%%%%%%%%%%%%%%%%%%%%%%%%%%%%%%%%%%%%%%
%\begin{sidewaystable}
\begin{table}[htbp]
\begin{center}
\caption{\label{tab:molecules-num-parameter} The numbers of valence electrons and variational parameters for the 8 hydrocarbons (CH$_4$, C$_2$H$_4$, C$_2$H$_6$, C$_6$H$_6$, C$_{10}$H$_8$, C$_{14}$H$_{10}$, C$_{18}$H$_{12}$, C$_{20}$H$_{10}$) and the C$_{60}$ fullerene. The numbers are plotted in the Fig.~4 of the main body.}
\begin{tabular}{c|c|ccc}
\Hline
         Formula &  Num. valence electrons &  Num. param. (JSD) &  Num. param. (JAGPn) &  Num. param. (JAGP) \\
\Hline
          CH$_4$ &                       8 &                 21 &                   25 &                1651 \\
      C$_2$H$_4$ &                      12 &                 27 &                   35 &                3729 \\
      C$_2$H$_6$ &                      14 &                 41 &                   40 &               10841 \\
      C$_6$H$_6$ &                      30 &                 57 &                   86 &               17629 \\
   C$_{10}$H$_8$ &                      48 &                 78 &                  135 &               41568 \\
C$_{14}$H$_{10}$ &                      66 &                194 &                  187 &              151066 \\
C$_{18}$H$_{12}$ &                      84 &                245 &                  234 &              239987 \\
C$_{20}$H$_{10}$ &                      90 &                332 &                  251 &              477318 \\
        C$_{60}$ &                     240 &                422 &                  647 &             1314634 \\
\Hline
\end{tabular}
\end{center}
\end{table}
%\end{sidewaystable}
%%%%%%%%%%%%%%%%%%%%%%%%%%%%%%%%%%%%%%%%%%%%%%%%%%%%

%%%%%%%%%%%%%%%%%%%%%%%%%%%%%%%%%%%%%%%%%%%%%%%%%%%%
%%%%%%%%%%%%%%%%%%%%%%%%%%%%%%%%%%%%%%%%%%%%%%%%%%%%

\subsection{Total energies of methane, water, and methane-water complexes}

\vspace{2mm}
Table~{\ref{tab:molecules-energy-water-methane}} shows the LRDMC total energies of methane, water, and methane-water complex. Hereafter, we discuss the role of molecular orbitals (MOs) or natural orbitals (NOs) in the AGPn ansatz. The comparison of the JSD energy with the HF orbitals and that with LDA orbitals reveals that the nodal surface obtained by LDA is better than that obtained by HF. This is also true in the JAGPn ansatz. The comparison between JAGPn with the LDA(HF) orbitals and JAGPn with LDA(HF)-MP2 orbitals tells us the importance of MOs or NOs employed in the expansion of the JAGPn ansatz. Indeed, the total energies obtained with the JAGPn consisting of the LDA(HF)-MP2 orbitals are lower than those with the JAGPn consisting of the LDA(HF) orbitals for the three molecules. It indicates that the NOs made from the MP2 calculations are better than the as-is LDA(HF) MOs in the expansion. This is because the MP2 virtual orbitals have more physical meanings than the LDA(HF) ones. The table also indicates that the NOs composed of the LDA orbitals are better than those composed of the HF orbitals. Thus, we concluded that the best strategy is making the JAGPn ansatz using the NOs composed of the LDA-MP2 orbitals, which are employed in the calculations reported in the main text.
Table~{\ref{tab:molecules-energy-water-methane}} also contains the results of the binding energy calculations for the methane--water dimer.

\vspace{2mm}
Table~{\ref{tab:water-mathane-faraway-energy}} contains the total energies of the faraway water-methane complex, $E_{\rm faraway}$, at a distance of $\sim 11$~\AA\ and the sum of the isolated molecules, $E_{\rm isolated}$, and the difference between them, $E_{\text{SCE}} \equiv E_{\rm isolated} - E_{\rm faraway}$. They were computed with the JSD ansatz, with the JAGPn ansatz optimized using either VMC or FN gradients, and with the JAGP ansatz optimised with VMC gradients.

%%%%%%%%%%%%%%%%%%%%%%%%%%%%%%%%%%%%%%%%%%%%%%%%%%%%
\begin{center}
\begin{table*}[htbp]
\caption{\label{tab:molecules-energy-water-methane} The total energies obtained by LRDMC ($a \rightarrow 0$) calculations for the methane, water, and the methane-water dimer.}
\vspace{2mm}
\begin{tabular}{ccc|ccc|c}
\Hline
Ansatz &      MO &   Opt. & CH$_{4}$ (Ha) & H$_{2}$O (Ha) & CH$_{4}$-H$_{2}$O (Ha) & Binding energy (meV) \\
\Hline
   JSD &      HF &      - &   -8.07801(7) &  -17.23413(8) &           -25.31290(8) &               -21(4) \\
 JAGPn &      HF &  FNopt &   -8.07821(7) &  -17.23484(7) &           -25.31373(7) &               -19(3) \\
 JAGPn &  HF-MP2 & VMCopt &   -8.07820(8) &  -17.23593(8) &           -25.31605(7) &               -52(4) \\
 JAGPn &  HF-MP2 &  FNopt &   -8.07864(7) &  -17.23630(8) &           -25.31580(7) &               -23(3) \\
\Hline
   JSD &     LDA &      - &   -8.07858(3) &  -17.23489(3) &           -25.31445(4) &               -27(2) \\
 JAGPn &     LDA &  FNopt &   -8.07909(3) &  -17.23594(3) &           -25.31600(4) &               -27(2) \\
 JAGPn & LDA-MP2 & VMCopt &   -8.07899(3) &  -17.23693(3) &           -25.31760(5) &               -46(2) \\
 JAGPn & LDA-MP2 &  FNopt &   -8.07940(3) &  -17.23718(3) &           -25.31765(7) &               -29(2) \\
\Hline
  JAGP &      NA & VMCopt &   -8.07902(3) &  -17.23736(8) &           -25.31789(8) &               -41(3) \\
\Hline
\end{tabular}
\end{table*}
\end{center}
%%%%%%%%%%%%%%%%%%%%%%%%%%%%%%%%%%%%%%%%%%%%%%%%%%%%

%%%%%%%%%%%%%%%%%%%%%%%%%%%%%%%%%%%%%%%%%%%%%%%%%%%%
% Comparison of Water-Methane energies
%%%%%%%%%%%%%%%%%%%%%%%%%%%%%%%%%%%%%%%%%%%%%%%%%%%%
\begin{center}
\begin{table*}[htbp]
\caption{\label{tab:water-mathane-faraway-energy} Comparison of LRDMC energies (LRDMC, $a \rightarrow 0$) of the far-away water-methane complex and the sum of the isolated water and methane molecules, obtained with various ansatz. $E_{\text{SCE}} = E_{\rm isolated} - E_{\rm faraway}$.}
\vspace{2mm}
\begin{tabular}{c|c|cc|c}
\Hline
Ansatz &   Opt. & E$_{\rm isolated}$ (Ha) & E$_{\rm faraway}$ (Ha) & $E_{\text{SCE}}$ (meV) \\
\Hline
   JSD &     -  &            -25.31347(4) &           -25.31344(3) &           -1(1) \\
 JAGPn & VMCopt &            -25.31592(4) &           -25.31630(7) &           10(2) \\
 JAGPn &  FNopt &            -25.31658(4) &           -25.31650(7) &           -2(2) \\
  JAGP & VMCopt &            -25.31638(8) &           -25.31679(7) &           11(3) \\
\hline
\Hline 
\end{tabular}
\end{table*}
\end{center}
%%%%%%%%%%%%%%%%%%%%%%%%%%%%%%%%%%%%%%%%%%%%%%%%%%%%

\vspace{2mm}
\subsection{Torsion energy of ethylene}
\begin{figure}[htbp]
  \centering
  \includegraphics[width=0.6\columnwidth]{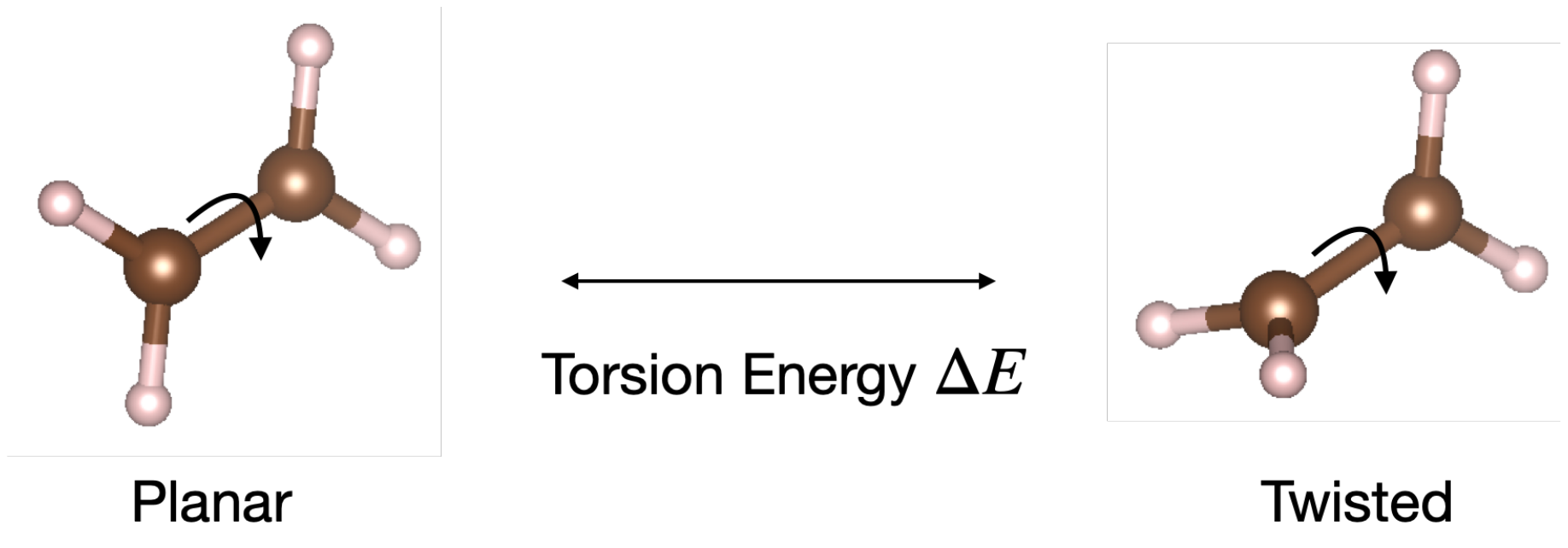}
  \caption{The schematic figure of the torsion between the planar and twisted ethylene.}
  \label{fig:ethylene-torsion}
\end{figure}

Figure.~{\ref{fig:ethylene-torsion}} shows the schematic figure of the ethylene torsion. The torsion energy is defined as the energy difference between the ground state ethylene (denoted as planer ethylene) and the orthogonally rotated ethylene (denoted as twisted ethylene). Table~{\ref{tab:ethylene-total-energy}} shows the total energies of the ethylenes computed with the JSD and JAGPn ansatz. Table~{\ref{tab:ethylene-torsion-energy}} summarizes the obtained torsion energies and reference values obtained in previous works. 

%%%%%%%%%%%%%%%%%%%%%%%%%%%%%%%%%%%%%%%%%%%%%%%%%%%%
%%%%%%%%%%%%%%%%%%%%%%%%%%%%%%%%%%%%%%%%%%%%%%%%%%%%

%%%%%%%%%%%%%%%%%%%%%%%%%%%%%%%%%%%%%%%%%%%%%%%%%%%%
% Total energies of Ethylene
%%%%%%%%%%%%%%%%%%%%%%%%%%%%%%%%%%%%%%%%%%%%%%%%%%%%
\begin{center}
\begin{table}[htbp]
\caption{\label{tab:ethylene-total-energy} Comparison of LRDMC energies ($a \rightarrow 0$) of the planar and twisted ethylene.}
\vspace{2mm}
\begin{tabular}{c|cc|c}
\Hline

        molecule &    JHF (Ha) & JAGPn-HF (Ha) & $\Delta E$ (mHa) \\
\Hline
 Planar Ethylene & -13.7099(5) &   -13.7106(5) &          -0.8(7) \\
Twisted Ethylene & -13.4977(5) &   -13.5943(4) &         -96.6(6) \\
\Hline
\end{tabular}
\end{table}
\end{center}
%%%%%%%%%%%%%%%%%%%%%%%%%%%%%%%%%%%%%%%%%%%%%%%%%%%%
% Torsion energy of Ethylene
%%%%%%%%%%%%%%%%%%%%%%%%%%%%%%%%%%%%%%%%%%%%%%%%%%%%
\begin{center}
\begin{table}[htbp]
\caption{\label{tab:ethylene-torsion-energy} The torsion energies between the planar and twisted ethylene, obtained with various approaches.}
\vspace{2mm}
\begin{tabular}{c|cc|c}
\Hline
      Approach & $\Delta E$ (kcal/mol) \\
\Hline
     LRDMC/JHF &                       133.1(4) \\
LRDMC/JAGPn-HF &                        73.0(4) \\
    LRDMC/JAGP{\footnotemark[1]} &                        70.2(2) \\
     MR-CISD+Q{\footnotemark[2]} &                        69.2    \\
\Hline
\end{tabular}
\footnotetext[1]{This value is taken from Ref.~{\onlinecite{2014ZEN}}.}
\footnotetext[2]{This value is taken from Ref.~{\onlinecite{2004BAR}}.}
\end{table}
\end{center}
%%%%%%%%%%%%%%%%%%%%%%%%%%%%%%%%%%%%%%%%%%%%%%%%%%%%

%%%%%%%%%%%%%%%%%%%%%%%%%%%%%%%
\bibliographystyle{apsrev4-1}
\bibliography{./references.bib}
%%%%%%%%%%%%%%%%%%%%%%%%%%%%%%%